\documentclass[twocolumn]{mn2e}
\usepackage{mnras_cite}
\usepackage{float}
\usepackage[dvips]{graphicx}
\usepackage{amsmath}

\newcommand{\beq}{\begin{equation}}
\newcommand{\eeq}{\end{equation}}

\newcommand{\beqa}{\begin{eqnarray}}
\newcommand{\eeqa}{\end{eqnarray}}

\newcommand{\nc}{\newcommand}   
\nc{\citealt}{\cite}

\def\Kpc{\, h^{-1} \, {\rm Kpc}}
\def\Mpc{\, h^{-1} \, {\rm Mpc}}
\def\Gpc{\, h^{-1} \, {\rm Gpc}}
\def\Mpccube{\, h^{-3} \, {\rm Mpc}^3}
\def\Gpccube{\, h^{-3} \, {\rm Gpc}^3}
\def\kvecMpc{\, h \, {\rm Mpc}^{-1}}
\def\Msun{\,h^{-1}\,{\rm M_{\odot}}}
\def\ltsima{$\; \buildrel < \over \sim \;$}   
\def\gtsima{$\; \buildrel > \over \sim \;$}   
\def\simlt{\lower.5ex\hbox{\ltsima}}   
\def\simgt{\lower.5ex\hbox{\gtsima}}   
\def\etal{{et al. }}

\input epsf

\begin{document}

\title[The abundance of massive clusters]
{Simulating the Universe with MICE : The abundance of massive clusters}

\author[Crocce \etal]{Mart\'in Crocce,
Pablo Fosalba,
Francisco J. Castander \&
Enrique Gazta\~{n}aga \\
Institut de Ci\`encies de l'Espai, IEEC-CSIC, Campus UAB, Facultat de Ci\`encies, Torre C5 par-2, Barcelona 08193, Spain}

\maketitle

\begin{abstract}
We introduce a new set of large N-body runs, the MICE simulations,
that provide a unique combination of very large cosmological volumes
with good mass resolution.
They follow the gravitational evolution of $\sim 8.5$ billion particles
($2048^3$) in volumes covering up to $\sim 15$ Hubble volumes
(i.e., $450\Gpccube$), and sample over $5$ decades in spatial resolution.
Our main goal is to accurately model and calibrate basic
cosmological probes that will be used by upcoming astronomical surveys of unprecedented volume.
Here we take advantage of the very large volumes of MICE to
make a robust sampling of the high-mass tail of the halo mass function (MF).
We discuss and avoid possible systematic effects in our study, and do a detailed analysis
of different error estimators.
We find that available fits to the local abundance of halos
(Warren et al. (2006)) match well the abundance estimated in the large volume of MICE  
up to $M\sim 10^{14}\Msun$, but significantly deviate for larger masses,
underestimating the mass function by $10\%$ ($30\%$) at $M =  3.16 \times 10^{14}\Msun$ ($10^{15}\Msun$). 
Similarly, the widely used Sheth \& Tormen (1999) fit,
if extrapolated to high redshift assuming universality,
leads to an underestimation of the cluster abundance
by $30\%$, $20\%$ and $15\%$ at $z=0$, $0.5$, $1$
for fixed $\nu = \delta_c/\sigma \approx 3$ (corresponding to
$M\sim [7 - 2.5 - 0.8] \times 10^{14}\Msun$
respectively).
We provide a re-calibration of the MF over $5$ orders of magnitude in mass
($10^{10} < M/(\Msun) < 10^{15}$), that accurately describes its redshift evolution up to $z=1$.
We explore the impact of this re-calibration
on the determination of the dark-energy equation of state $w$, and conclude
that using available fits that assume universal evolution for the cluster MF
may systematically bias the estimate of $w$ by as much as
$50\%$ for medium-depth ($z \simlt 1$) surveys.
The halo catalogues used in this analysis are
publicly available at the MICE webpage, ${\rm http://www.ice.cat/mice}$.
\end{abstract}

\maketitle

\section{Introduction}
\label{sec:introduction}

Near future extra-galactic surveys will sample unprecedentedly large cosmological volumes, 
in the order of tens of cubic gigaparsecs, by combining wide fields with deep spectroscopy or 
photometry, typically reaching $z \sim 1$ (e.g. DES, PAU, BOSS, PanSTARRS, WiggleZ 
\footnote{
DES (http://www.darkenergysurvey.org/);      \newline 
PAU (http://www.ice.csic.es/pau/Survey.html); \newline
BOSS (http://cosmology.lbl.gov/BOSS/);        \newline
PanSTARRS (http://pan-starrs.ifa.hawaii.edu); \newline
WiggleZ (http://wigglez.swin.edu.au/)
}). 
In addition they will be able to capture very faint objects and lower their shot-noise level 
to become close to sampling variance limited. 
Optimizing the preparation and scientific exploitation of these upcoming large surveys requires
accurate modeling and simulation of the expected huge volume of high quality data. 
This is quite a non-trivial task, because it involves simulating a wide dynamic 
range of cosmological distances in order to accurately sample the large-scale structure, 
and well resolved dark-matter halos as a proxy to
galaxy clusters, to model the physics of galaxy formation and other nonlinear physics. 

Over the past decades numerical simulations have provided one of the most valuable tools to address these issues, and their relevance will certainly increase in the near future.  They allow to follow the growth of cosmological structure, shed light on the process of galaxy formation, model non-linear effects entering different clustering measures, lensing and redshift distortions, track the impact of a dark-energy component and more. Among projects related to the development of very large-simulations are those carried out by the Virgo consortium \cite{virgo}, the Millennium I and II \cite{millenium,millenium2}, the Horizon run  \cite{horizonrun} and the Horizon project \cite{horizon}. 
They have all benefited by the vast computational power and hardware developed over the past years. 

In this paper we present a new effort to tackle the demand of large simulations 
and mock catalogues, the MareNostrum Institut de Ci\`encies de l'Espai (MICE) simulations, 
that aims at the development of very large and comprehensive N-body runs to deliver an unprecedented
combination of large simulated volumes with good mass resolution.

As a first step, we have developed two N-Body simulations including more than 8 billion particles 
($2048^3$) each, in volumes similar and well beyond the one corresponding to the Hubble length ($\sim 30 \Gpccube$), in addition to several other large runs of typically smaller volume and corresponding 
higher mass resolution that are complementary to the large volume runs.

Some of the MICE simulations used in this paper have already been used 
to develop the first full-sky weak-lensing maps in the lightcone \cite{fosalba08}, or study the clustering of 
LRG galaxies with multiple-band photometric surveys such as PAU \cite{pau}. More recently, 
using the largest volume simulations, a series of papers has studied the large-scale clustering in the 
spectroscopic LRG SDSS sample, through the redshift space distortions \cite{cabre09a,cabre09b}, 
the baryon oscillations in the 3-point function \cite{gazta08a}, and in the radial direction \cite{gazta08b}.

In this paper we will focus on the mass function of the most massive objects formed through hierarchical clustering, since their 
low abundance makes the need of large sampling regions crucial. In turn, a precise description of this regime is of paramount 
importance since the abundance of clusters, to which it corresponds to, is very sensitive to cosmological parameters (particularly
 the matter density), the normalization of the matter power spectrum and the expansion history of the universe, characterized by the 
dark energy density and its equation of state (e.g. see \pcite{rozo09,cunha2009,mantz08,henry09,vikhlinin09} and references therein). 
This regime is also one of the best probes to search for  primordial non-Gaussianities (e.g. \pcite{matarrese00,grossi07,pillepich08,maggiore09})

The halo mass function and related topics have been extensively studied in the literature. Analytic models predicting not only the abundance 
as a function of mass but also the evolution were developed as far back as the $70'$s by \cite{PS} and followed by \cite{bond91,lee98,sheth01}. 
However the development of N-body simulations showed that these predictions were in general not sufficiently accurate for cosmological 
applications, and demanded the need for calibrations against numerical results (see also \pcite{robertson09}). 
Although N-body studies were likewise early developed (e.g. two $1000$ particles simulation by \pcite{PS}), the reference work in this 
directions was set by \cite{ST99} and \cite{HVS}.
More recent re-calibrations of the mass function to within few percent were put forward by \cite{warren06,tinker08}. In addition these
 or other papers have focused their attention on the redshift evolution of the mass function \cite{reed03,reed07,lukic07,cohn07}, 
different definitions of halo and halo mass \cite{white01,white02,tinker08} or the impact of gas physics associated with halo
 baryons \cite{stanek09baryon}.

In this paper we combine the effect of long-wavelength modes whose contribution can only be studied with the unprecedented volume 
of the MICE simulations ($\sim 30, 100$ and $450 \Gpccube$), with good mass resolution and controlled systematics to investigate how 
well available fits describe the high-mass end tail of the halo mass function. We complement this with a nested-volume approach of 
N-body runs to probe smaller masses in a way to sample the mass function over $5$ decades in mass.

Before proceeding we recall that in hierarchical models of structure formation, as our present 
cosmological paradigm, halos are often found merging or accreting rather than isolated, and therefore have no definite boundary. 
Hence defining a halo and its mass becomes rather conventional.

The are two widely used conventions. The spherical overdensity (SO, \pcite{lacey94}) halos are defined as spherical regions around matter 
density peaks with an inner density larger than a given threshold, which is generally taken as a fixed multiple of the critical or background 
density. Alternatively, the Friends-of-Friends (FoF, \pcite{fof}) algorithm identifies all the neighbors to a given particle separated by less
 than a fixed distance (the linking length). The same algorithm is then applied to each neighbor recursively until no more ``friends'' 
are found. The end result is a group of particles in space whose boundary approximately matches an iso-density contour 
(e.g. see Fig 1 in \pcite{lukic09}).

Each definition has advantages and disadvantages, and the overall convenience of one over the other is currently an open debate 
\cite{HVS,white02,tinker08,lukic09}. The abundance of FoF halos has been found universal at $< 10\%$ level by different studies 
including ourselves in Sec.~\ref{sec:MF_growth} \cite{HVS,reed03,heitmann06,reed07,lukic07,tinker08}, with some dependence on the 
linking length \cite{HVS,tinker08}. Given the complicated non-linear processes that drive halo formation, having a universal mass-function 
(i.e. independent of red-shift and cosmology) at this level could be very convenient alleviating the need to simulate individual cosmologies.
 In contrast, the abundance of SO halos is considerably less universal, showing evolution in the halo mass function amplitude at the 
level of $20\%-50\%$, depending on cosmology, but also evidence for red-shift dependence of its overall shape \cite{tinker08}.
The current error budget of cluster science is dominated by the unknowns in the calibration of mean and scatter of the mass-observable
 relations (e.g. \pcite{voit05,rozo09,stanek09}). Only after understanding these better we will be able to state
what is an acceptable degree of non-universality in future precision cosmology (but see also \pcite{stanek09baryon}).

In turn, the general idea of defining virialized structure and mass in terms of spherical apertures is more directly linked to observations 
of galaxy groups and clusters (e.g. \pcite{voit05}), in part because the scaling correlations between cluster observable and mass are tighter 
in this approach. Nonetheless, relating iso-density methods such as FoF is also possible (e.g \pcite{eke06,tago08,wen09}). 

For concreteness, we will next focus on FoF halos and leave SO for a follow-up study, although we have tested to 
some extent that our main results are robust in front of this choice, as we further discuss in our Conclusions.

This paper is organized as follows: In Sec.~\ref{sec:MICE} we describe the MICE simulations. Sec.~\ref{sec:massfunc} recaps 
known theoretical predictions and fits to the halo mass function and 
concludes with a comparison between MICE and results from previous simulations. 
In Sec.~\ref{sec:systematics} we discuss systematic effects 
that are most relevant in the measurement of the high-mass end of the mass function, 
such as transients from initial conditions, finite sampling of the mass distribution, and mass resolution effects. 
A detailed error analysis including different estimators is provided in Sec.~\ref{sec:Errors}, 
whereas in Sec.~\ref{sec:fit} we derive a new fitting function to account for the high-mass tail 
of the halo mass function. The higher redshift evolution, including results regarding mass function universality, 
is the subject of Sec.~\ref{sec:MF_growth}. 
We discuss the implications of our results for dark-energy constraints in Sec.\ref{sec:cosmo}, 
and we finish by summarizing and discussing our main findings in Sec.~\ref{sec:Conclusions}.

\section{The MICE simulations}
\label{sec:MICE}

The set of large N-body simulations described in this paper were carried out on the Marenostrum supercomputer at the Barcelona Supercomputing Center (http://www.bsc.es), hence their acronym MICE (Marenostrum-Instituto de Ciencias del Espacio). 

All simulations were ran with the Gadget-2 code \cite{gadget2} assuming the same flat concordance LCDM model with parameters $\Omega_m=0.25$, $\Omega_{\Lambda}=0.75$, $\Omega_b=0.044$, and $h=0.7$. The linear power spectrum had spectral tilt $n_s=0.95$ and was normalized to yield $\sigma_8=0.8$ at $z=0$. Special care was taken in order to avoid spurious artifacts from the initial conditions ({\it transients}).
Thus, the initial particle distributions were laid down either using the Zeldovich approximation with a high starting redshift or 2$nd$ order Lagrangian Perturbation Theory (2LPT) \cite{2LPTorig,2LPT} (see Sec.~\ref{Sec:transients} for details). 

\begin{figure}
\begin{center}
\includegraphics[width=0.5\textwidth]{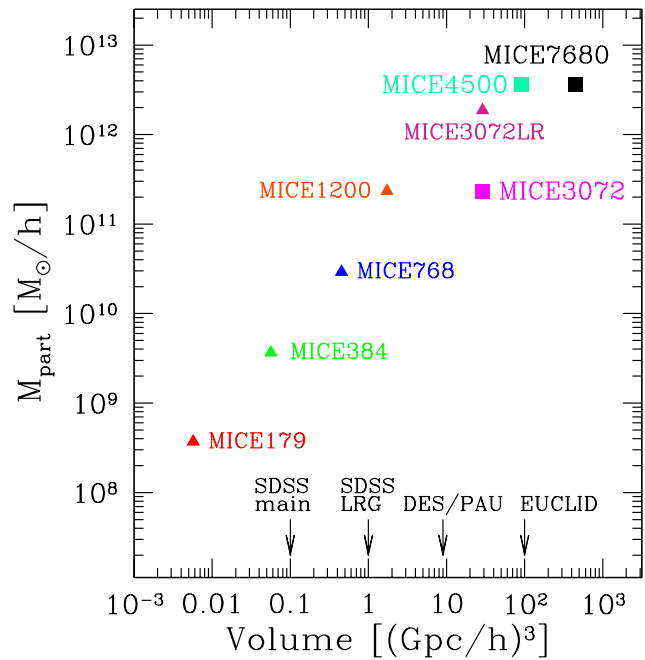}
\caption{{\it The MICE simulations in the mass resolution-volume plane:} they 
span over volumes comparable to the SDSS-main sample ($0.1 \Gpccube$), SDSS-LRG survey ($1 \Gpccube$), 
DES or PAU surveys ($9 \Gpccube$), and up to huge volumes such as the planned EUCLID mission ($100 \Gpccube$), 
and deliver mass resolutions from $3\times 10^{12}\Msun$ down to $3\times 10^8 \Msun$. In turn, the largest volume simulations (big squares) map the mass function at the high-mass end, $\sim 10^{15} \Msun$, whereas the test simulations (small triangles) extend the dynamic range down to halos of $10^{10} \Msun$.} 

\label{fig:micesims}
\end{center}
\end{figure}

The main goal of the MICE set is to study the formation and evolution of structure at very large scales, with the aim of simulating with enough mass resolution the size of future large extra-galactic surveys, such as DES \cite{des} or PAU \cite{pau}, and test robustly statistical and possible systematic errors.
Fig.\ref{fig:micesims} shows the set of MICE simulations in the mass resolution-volume plane. They 
sample cosmological volumes comparable to the SDSS main sample ($0.1 \Gpccube$), the SDSS-LRG survey ($1 \Gpccube$), 
PAU or DES ($9 \Gpccube$), and those of huge future surveys such as EUCLID ($\sim 100 \Gpccube$), 
in combination with mass resolutions from $3\times 10^{12}\Msun$ down to $3\times 10^8 \Msun$. 
In turn, the largest volume simulations (squares) map the mass function at the high-mass end, $\sim 10^{15} \Msun$, 
whereas the test simulations (triangles) extend the dynamic range down to halos of $10^{10} \Msun$.
Table \ref{simtab} summarizes the identifying parameters of the main MICE simulations.


\begin{table*} 
\begin{center}
\begin{tabular}{lcccccccccccc}
\hline \\
Run        &&    $N_{{\rm part}}$ && \ $L_{{\rm box}}/\Mpc$  \ && \  $m_p/\Msun$ && $l_{{\rm soft}}/\Kpc$ && IC  && $z_{{\rm i}}$   \\
           &&              &&               &&                   &&        &&     &&     \\
MICE7680   &&    $2048^3$  && $7680$        && $3.66   \times 10^{12}$  && 50 && ZA  && 150      \\
MICE3072   &&    $2048^3$  && $3072$        && $2.34   \times 10^{11}$ &&  50 && ZA  && 50        \\ 
MICE4500   &&    $1200^3$  && $4500$        && $3.66   \times 10^{12}$   && 100 && 2LPT && 50     \\  \\
MICE3072LR$^{\star}$ &&    $1024^3$  && $3072$        && $1.87   \times 10^{12}$ && 50 && ZA && 50        \\ 
MICE768$^{\star}$    &&    $1024^3$  && $768$         && $2.93   \times 10^{10}$  && 50 && 2LPT && 50      \\
MICE384$^{\star}$    &&    $1024^3$  && $384$         && $3.66   \times 10^{9}$  && 50 && 2LPT && 50      \\
MICE179$^{\star}$    &&    $1024^3$  && $179$         && $3.70   \times 10^{8}$  &&  50 && 2LPT && 50     \\
MICE1200$^{\star}$ ($\times 20$)   &&   $800^3$ && $1200$ && $2.34 \times 10^{11}$  &&  50 && ZA && 50 \\  \\
\hline
\end{tabular}
\end{center}
\caption{Description of the MICE N-body simulations. $N_{{\rm part}}$ denotes number of particles, $L_{{\rm box}}$ is the box-size, $m_p$ gives the particle mass, $l_{soft}$ is the softening length, $IC$ is the type of initial conditions (Zeldovich Approximation, ZA, or 2nd order Lagrangian Perturbation Theory, 2LPT), 
and  $z_{in}$ is the initial redshift of the simulation.
Their cosmological parameters were kept constant throughout the runs (see text for details), the initial global time-step is of order $1\%$ of the Hubble time (i.e, $d \log a=0.01$, being $a$ the scale factor), and the number of global timesteps to complete the run $N_{steps} \simgt 2000$ in all cases. We ran an ensemble of 20 different realizations with the parameters of MICE1200 primarily to calibrate error estimators. We mark with $^\star$ those runs that were done for completeness or testing as main purpose.}
\label{simtab}
\end{table*}


Notice that for one particular case (MICE1200) we implemented a set of $20$ independent realizations, in order to compare statistical errors on different quantities obtained from an strictly ``ensemble error'' approach from other internal or external error estimates.

In addition to the production of comoving outputs at several redshifts, we have constructed projected density and weak lensing maps as well as ligth-cone outputs from the main MICE runs. The mass projected and lensing catalogues were discussed in \cite{fosalba08}, while the light-cone catalogue will be presented in future work. Note that both represent a huge compression factor ($\sim 1000$), that may turn essential in dealing with very large number of particles as in our case. Further details and publicly available data can be found at http://www.ice.cat/mice.

\section{The Halo Mass Function}
\label{sec:massfunc} 

The very large simulated volume spanned by the MICE set allow us to study accurately not only Milky Way size halos, but specially the most massive and rarest halos formed by hierarchical clustering. To this end we built dark matter halo catalogues at each snapshot of interest according to the Friends-of-Friends (FoF) algorithm \cite{fof} with linking length parameter $b$ set in units of the mean inter-particle distance in each simulation. We will refer to halos defined in this way as FoF(b). For the most part we will deal with the $b=0.2$ catalogues, although we have also implemented $b=0.164$ for a first validation of our simulations against the Hubble Volume Simulation (HVS) \cite{HVS,HVS2}. The HVS is one of the very few publicly available halo catalogue comparable in simulated volume to MICE.

The halo finder algorithm was implemented using the FOF code publicly available at the $N$-body Shop (http://www-hpcc.astro.washington.edu/), with some additional modifications needed in order to handle the large number of particles in reasonable amount of time. The resulting halo catalogues contain not only the mass, position and velocity of the center of mass, and virial velocity, but also information of all the particles forming each halo.

As an example of the size of our outputs we mention that MICE3072 contains at $z=0$ a total of about $25$ million halos more massive than $3.9\times 10^{12}\Msun$ if the minimum number of particles per halo is set to $20$. The most massive object weighs $5.27\times 10^{15}\Msun$ and is made of $22,561$ particles. In turn, MICE7680 contains about $15$ million halos with mass greater than $7.3\times 10^{13} \Msun$, with the biggest reaching $8.4\times 10^{15} \Msun$.

\subsection{Theoretical Predictions}

Let us start by recalling some well known results regarding the abundance of halos. The differential mass function is defined as,
\begin{equation}
f(\sigma,z)=\frac{M}{\rho_b} \frac{dn(M,z)}{d\ln{\sigma^{-1}(M,z)}}
\label{fnu}
\end{equation}
where $n(M,z)$ is the comoving number density of halos with mass $M$ and $\sigma(M,z)$ is the variance of the linear density field smoothed with a top hat filter of radius $R$ and enclosing an average mass $M=\rho_b 4\pi R^3 /3$,
\beq
\sigma^2(M,z)=\frac{D^2(z)}{2\pi^2} \int k^2 P(k) W^2(k R) dk ,
\label{eq:sigma8}
\eeq
with
\beq
W(x)=\frac{3}{(x)^3}[\sin(x)-x\cos(x)]. \nonumber
\label{eq:window}
\eeq
In Eq.~(\ref{eq:sigma8}), $D(z)$ is the linear growth factor between $z=0$ and the redshift of interest, 
and $P(k)$ the linear power spectrum of fluctuations at $z=0$.

In Eq.~(\ref{fnu}) we have explicitly assumed that all the cosmology dependence of the differential mass function enters through the amplitude of linear fluctuations, Eq.~(\ref{eq:sigma8}), at the mass scale $M$. If the redshift dependence also satisfies this condition the halo abundance as defined by Eq.~(\ref{fnu}) is said to be universal \cite{PS,ST99,HVS}.

Several analytical derivations \cite{PS,bond91,sheth01} or fits \cite{ST99,HVS,white02,reed03,reed07,warren06,tinker08} have been provided in the literature over the past years, starting from the original Press-Schechter formalism in 1974 \cite{PS}. In our work we will refer only to the Sheth and Tormen (ST) fit given by \cite{ST99},
\begin{equation}
f_{\rm ST}(\sigma)= A \sqrt{\frac{2 q}{\pi}} \frac{\delta_c}{\sigma} \left[1+\left(\frac{\sigma^2}{q \delta_c^2}\right)^p   \right] \exp\left[-\frac{q \delta_c^2}{2 \sigma^2}\right],
\label{eq:ST}
\end{equation}
with $A=0.3222$, $q=0.707$ and $p=0.3$. 
In addition we will take the value of the linear over-density at collapse as $\delta_c=1.686$, and ignore its weak dependence on cosmology \cite{lacey93,HVS}. The subsequent Jenkins fit \cite{HVS},
\begin{equation}
f_{\rm Jenkins}(\sigma)= A \exp\left[-|\log \sigma^{-1}+b |^{c}\right],
\label{eq:jenkins}
\end{equation}
with $A=0.315$, $b=0.61$ and $c=3.8$ corresponding to FoF(0.2) halos, that was obtained at redshifts $z=0-5$ over the range $-1.2\le \ln \sigma^{-1} \le 1.05$. For our cosmology this corresponds to masses $(0.96\times10^{10}-4.0\times10^{15})\Msun$. Alternatively, we will refer to his fit for FoF(0.164) halos for which $A=0.301$, $b=0.64$ and $c=3.88$ (Eq. B2 in \pcite{HVS}). This is valid over the mass range $(8.7\times10^{10}-3.4\times10^{15})\Msun$ in our cosmology.

More recently \pcite{warren06} performed a detailed mass function analysis using a set of nested-volume simulations and provided the following values for the best-fit parameters of a ST-like mass function,
\begin{equation}
f_{\rm Warren}(\sigma)= A \left[\sigma^{-a} +b\right] \exp\left[-\frac{c}{\sigma^2}\right],
\label{eq:warren}
\end{equation}
with $A=0.7234,a=1.625,b=0.2538,c=1.1982$, obtained from a fit to the mass range $(10^{10}-10^{15})\Msun$ at $z=0$. We will use Eq.~(\ref{eq:warren}) as our bench-mark reference fit.

\subsection{The binned Mass Function}
\label{sec:binmf}

In order to compare the predictions for the differential mass function in Eq.~(\ref{fnu}) with the observed halo abundance in a simulation of volume $L_{box}^3$ one would measure the number of halos $\Delta N$ in a given mass-bin $[M_1-M_2]$ of width $\Delta M$ and characteristic mass $M$,  and define,
\begin{equation}
\frac{dn}{d\ln M} = \frac{M}{L_{box}^3} \frac{\Delta N}{\Delta M}
\end{equation}
that is then related to the differential mass function in Eq.~(\ref{fnu}) after multiplying by the prefactor $-\sigma / \sigma^{\prime} \rho_b$. However for the most part we will directly compare the number density of halos in a given mass bin with the prediction binned in the same way as the measurements. That is, from Eq.~(\ref{fnu}) we obtain the theoretical number density of objects per unit mass $dn / dm$, which is then integrated as,
\beq
n_{bin} = \int_{M_1}^{M_2} \left(\frac{dn}{dm}\right) dM = \int_{M_1}^{{M_2}} \frac{-\rho_b}{M}\frac{1}{\sigma}\frac{d \sigma}{dM} f(M,z)\,dM 
\label{eq:n}
\eeq
to predict $\Delta N/V$. The corresponding value of the mass of the bin is obtained as
\beq
M_{bin} = \int_{M_{1}}^{M_{2}} \left(\frac{dn}{dm}\right) M dM
\label{eq:m}
\eeq
from the theory and as a mass weighted average, $M=\sum_{bin} M_i / \Delta N$, from the simulation. Throughout our study we used mass bins equally spaced in log-mass, with $\Delta \log_{10} M/(\Msun)  = 0.1$. We have tested that our conclusions do not depend on this particular choice.

\subsection{Comparison with previous work}

As a first validation of our set of large volume simulations we compared the halo abundance in MICE3072 to that in the Hubble Volume Simulation (HVS) \cite{HVS,HVS2}, since they both simulate almost the same total volume. The HVS is among the biggest simulated volume with halo data publicly available \footnote{We do not include a comparison to the recently available data from the Horizon run \cite{horizonrun} because of potential systematic issues in that simulation that are expected to affect the abundance of the most massive halos in a substantial manner, such as a low starting redshift $z_i=23$ combined with the ZA (as discussed in Sec.~\ref{Sec:transients}). Besides, this simulation uses a rather large force softening length $f_{\epsilon}=160\,{\it h}^{-1}\,{\rm kpc}$ and low number of global time-steps, $N_{steps}=400$.}, at http://www.mpa-garching.mpg.de/Virgo/hubble.html.

We used the catalogue corresponding to a $\Lambda$-CDM cosmology and FoF halos with linking length parameter $b=0.164$ \cite{HVS}. Thus, in what follows we will refer to the MICE catalogues for this value of $b$. Finally, for this comparison we employed the low resolution run of MICE3072 (MICE3072LR, see Table \ref{simtab}) that has a similar mass resolution to that in the HVS.

\begin{figure}
\begin{center}
\includegraphics[width=0.49\textwidth]{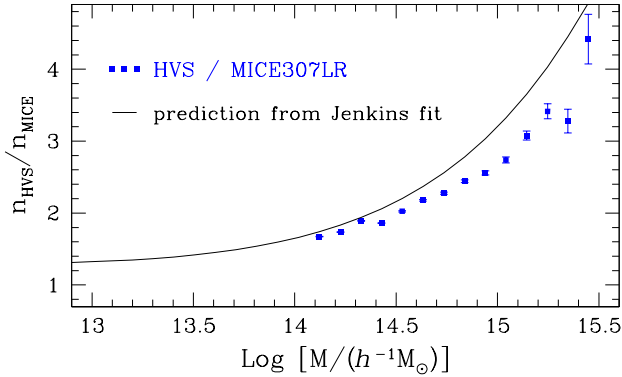}
\caption{{\it Comparison to the Hubble Volume Simulation (HVS):} 
We show results for the FoF halo mass functions at $z=0$ for a linking length parameter $b=0.164$
for the HVS in ratio with our low resolution run 
of MICE3072 (see the MICE3072LR entry in Table 1). We note that both runs
have a similar volume and particle mass, but they simulate different cosmologies 
(mainly a lower $\sigma_8$ for MICE3072LR). 
The solid line corresponds to the prediction for this ratio using the Jenkins fit for FoF(0.164) 
in Eq.~(\ref{eq:jenkins}). 
The difference between symbols and the prediction is within the accuracy 
claimed for the Jenkins fit. 
In all cases we show only halos with no less than 50 particles, and Poisson errors.}

\label{fig:MICEvsHVS}
\end{center}
\end{figure}

Figure~\ref{fig:MICEvsHVS} shows the ratio of the mass functions measured in the HVS and MICE3072LR at $z=0$. The higher abundance of massive halos in the HVS is due mainly to its larger value of $\sigma_8$, equal to $0.9$ against $0.8$ in MICE. Therefore we include in the figure the expected value for this ratio as predicted by the Jenkins fit in Eq.~(\ref{eq:jenkins}) (or Eq. B2 in \pcite{HVS}). The difference between the measured ratio and the prediction are within the claimed accuracy for the Jenkins fit ($10-15$ $\%$). Nonetheless notice that the HVS has a rather late start $z_i=34$ that leads to an artificially lower abundance (see \pcite{tinker08} for a discussion on this). If one corrects for this effect, the ratio of data-points in Fig.~\ref{fig:MICEvsHVS} increases by about $5\%$ at $10^{15}\Msun$ explaining part of the difference. We thus conclude that for matching volume (and particle mass) our MICE run agrees with the HVS.

\section{Systematic Effects}
\label{sec:systematics}

Measurements of the high-mass end of the halo mass function are potentially affected by a number of systematics.
Below we investigate in detail the most relevant ones: the impact of the choice of initial conditions, 
discrete sampling of the halo mass profile and mass resolution effects.

\subsection{Transients from initial conditions}
\label{Sec:transients}

Several potential sources of systematic errors must be considered and controlled when implementing an N-body run, with their relevance sometimes dictated by the regime at which one is interested (see \pcite{lukic07} for a detailed analysis). We have performed convergence test regarding force and mass resolution, initial time steeping, finite volume effects, and more. But of particular relevance to the abundance of the largest halos at a given output is the initial redshift and the approximate dynamics used to set the initial conditions and start the run (\pcite{2LPTorig,2LPT,tinker08,knebe09}). 

The generally adopted way to render the initial mass distribution is to displace the particles from a regular grid or a glass mesh, using the linear order solution to the equations of motion in Lagrangian Space. This is known as Zeldovich approximation (ZA). Particle trajectories within the ZA follow straight lines towards the regions of high initial overdensity. The ZA correctly describes the linear growth of density and velocity fields in Eulearian Space but, failing to account for tidal gravitational forces that bend trajectories, underestimates the formation of non-linear structure. To leading order this can be incorporated using the second order solution in Lagrangian Perturbation Theory (2LPT) effectively reducing the time it takes for the correct gravitational evolution to establish itself (known as {\it transients}) once the N-body started. During the period where transients are present the abundance of the most massive objects, that originate from the highest density peaks, is systematically underestimated.

In \pcite{2LPT} it is shown that transients affect the $z=0$ mass function reducing it by $5\%$ at $10^{15}\Msun$ if ZA, as opposed to 2LPT, is used to start at $z_i=49$. This value rises to $10\%$ for $M>2\times 10^{14}\Msun$ at $z=1$. Also \pcite{tinker08} finds evidence for transients in the HOT runs introduced in \pcite{warren06} and the HVS \cite{HVS}. These runs were started in the redshift range $z=24$ to $35$ using ZA. However their own run with $z_i=60$ is in good agreement with the 2LPT predicted abundance from \pcite{2LPT} by $z=1.25$. The impact of the starting redshift in the high redshift mass function has been investigated in \pcite{lukic07,reed07,reed03}.

\begin{figure}
\begin{center}
\includegraphics[width=0.45\textwidth]{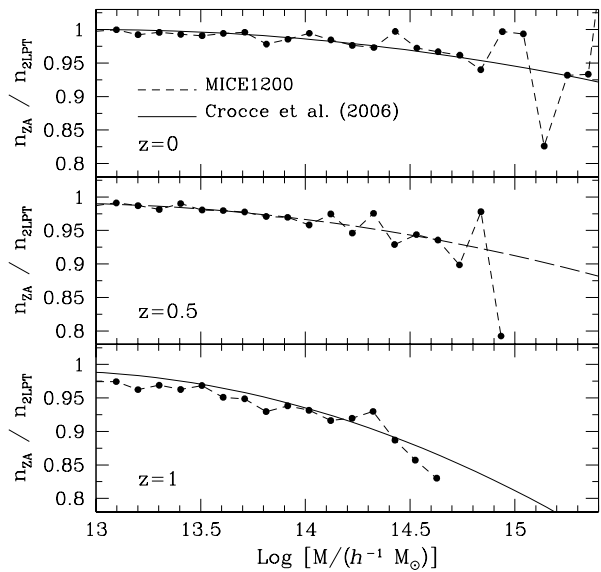}
\caption{{\it Transients in the mass function:} ratio of mass functions measured at $z = 0$ (top panel), $z =0.5$ (middle) and $z = 1$ (bottom), for N-body runs similar to MICE1200 (see Table \ref{simtab}) started at $z_i=50$ using either ZA or 2LPT to set-up the initial conditions. The solid line at $z=0,1$ panels displays the result of a comparable work done by Crocce \& Scoccimarro (2006), but for a different cosmology and mass resolution. Middle panel shows a 2nd order polynomial fit to the ratio $n_{\rm 2LPT}/n_{\rm ZA}$. Clearly, the approximate dynamics and starting redshift used to set-up initial condition plays a substantial role in the abundance of the rarest halos.}
\label{fig:transients}
\end{center}
\end{figure}

To test the significance of transients ourselves we implemented two runs of MICE1200 ($L_{\rm box}=1200\Mpc$ and $800^3$ particles) using ZA and 2LPT, both with $z_i=50$, and the same initial random phases (not listed in Table \ref{simtab}). Figure~\ref{fig:transients} shows the ratio of the measured mass functions, $n_{\rm ZA}/n_{\rm 2LPT}$, at $z=0$ (top panel), $z=0.5$ (middle) and $z=1$ (bottom). Top and bottom panels show in addition the results obtained by \pcite{2LPT} for the same combination of $[z,z_i]$ with a solid line. The dash line in the middle panel corresponds to a simple 2$nd$ order polynomial fit to the ratio at $z=0.5$. Our results agree very well with those in \pcite{2LPT} despite the difference in cosmology of the N-body runs (most notably $\sigma_8$), confirming an underestimation of the halo abundance by $\sim 5\%$ at $M \sim 10^{15},3.16\times 10^{14}$ and $10^{14}\Msun$  at $z=0,0.5$ and $1$ respectively (and larger for larger masses, at fixed redshift) if ZA $z_i=50$ is used instead of starting at higher redshift or using 2LPT. 

In line with the results above, almost all our runs in Table~\ref{simtab} were started  using 2LPT at $z_i=50$ to avoid transients in the low-redshift outputs. The convergence of 2LPT with $z_i \sim 50$ is discussed in detail in \pcite{2LPT}. For MICE7680 we implemented ZA at high starting redshift ($z_i=150$) to minimize transients. In this case the convergence is assured by the results in Fig.~\ref{fig:MF_z0}, where its halo abundance is compared with the one in MICE4500, that was started with 2LPT at $z_i=50$ with completely different random phases (particle load and volume are different). The measured abundance is practically indistinguishable. The only run expected to be affected by transients was  MICE3072 that uses ZA at $z_i=50$. In what follows we will therefore correct the mass function measured in MICE3072 by a simple fit to the ratios shown in Fig.~\ref{fig:transients}.

\subsection{FoF Mass Correction}
\label{sec:FoFmasscorrection}

As noted by \pcite{warren06} the mass of halos determined using the FoF algorithm suffer from a systematic over-estimation due the statistical noise associated with sampling the mass density field of each halo with a finite number of particles. By systematically sub-sampling an N-body simulation and studying the associated FoF(0.2) halo abundance (keeping the linking length parameter fixed) \pcite{warren06} determined an empirical correction of the mass bias that depends solely in the number of particles $n_p$ composing the halo through the simple expression,
\beq
n_p^{corr}=n_p (1-n_p^{-0.6}).
\label{eq:warrencorrection}
\eeq
However, as remarked by \pcite{lukic07}, the correction should be checked in a case-by-case basis since it is not the result of a general derivation (see also \pcite{tinker08}). While it is true that for well sampled halos the correction is relatively small (e.g.  $2.5\%$ for halos with $500$ particles) the impact that a few percent correction to the mass has in the halo abundance can be large is one refers to the most massive halos living in the rapidly changing high-mass tail of the mass function, as we are investigating in this paper. 

For this reason we have carried out an independent check of the correction in Eq.~(\ref{eq:warrencorrection}), with particular emphasis in the regime $M > 10^{13-14} \Msun$, where the mass function is exponentially suppressed. 

\begin{figure}
\begin{center}
\includegraphics[width=0.44\textwidth]{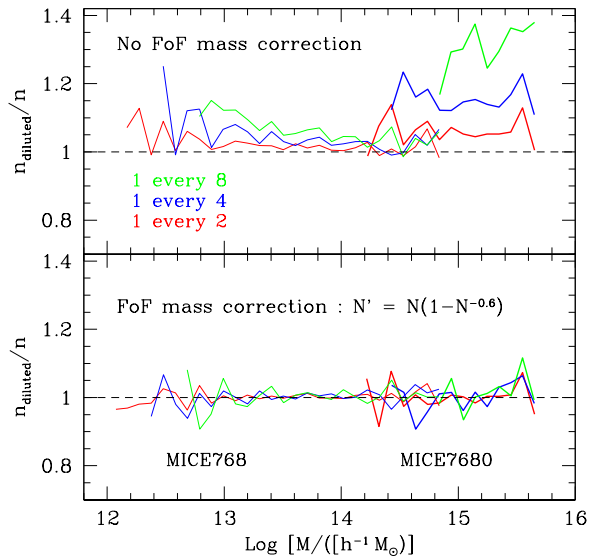}
\caption{{\it FoF Mass Correction:} We tested the correction to the systematic mass over-estimation intrinsic to the FoF algorithm (Warren et al. (2006)) in all of our dynamic range, but with particular emphasis at the very high-mass end (i.e. using MICE7680 and MICE4500). We randomly selected one every $n$ particles ($n=2,4,8$ in red, blue and green respectively) and ran the FoF algorithm afterwards with a linking length $n^{1/3}$ larger. The figure shows the corresponding ratio to the fully sampled mass function before (top panel) and after (bottom) the correction, for the cases of MICE7680 and 768. Notably, in all of our simulations the simple expression in Eq.~(\ref{eq:warrencorrection}) brings the full and sub-sampled mass functions into agreement.}
\label{fig:FoFmasscorrection}
\end{center}
\end{figure}

We randomly sub-sampled {\it every} simulation in the MICE set to several degrees (1 every $n=2, 4$ and $8$ particles) and run the FoF algorithm afterwards keeping the linking length parameter $b=0.2$ fixed (i.e. with the link length $n^{1/3}$ larger in each case). Results are shown in Fig.~\ref{fig:FoFmasscorrection} for two representative cases, MICE768 and MICE7680, but they extrapolate to all others in Table \ref{simtab}.

The correction in Eq.~(\ref{eq:warrencorrection}) is able to bring the sub-sampled mass functions into agreement with the original fully sampled one over the whole dynamic range (up to $4\times 10^{15} \Msun$). Most notably in the case of MICE7680 whose particle mass and volume makes it sample the mass function exponential tail with low Poisson shot-noise but with halos of no more than $\sim 2300$ particles, what makes it very sensible to such mass corrections. Finally, we have also tested that varying the factor $0.6$ leads to worse matching.

Hence, in what follows we will refer to the abundance of {\it mass corrected} FoF halos, unless otherwise stated.

\subsection{Mass Resolution Effects}
\label{sec:massres} 

\begin{figure}
\begin{center}
\includegraphics[width=0.49\textwidth]{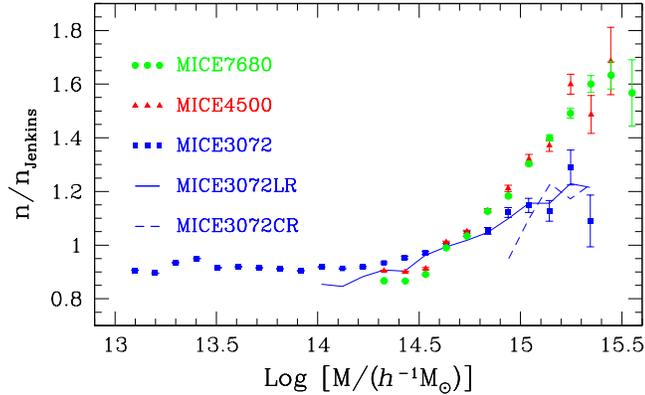}
\caption{{\it Mass Resolution effects on the high-mass end:} 
We show the abundance of FoF(0.164) halos at $z=0$ in MICE3072 in blue squares, MICE4500 in red triangles and MICE7680 in green circles. We find a systematic rise over the Jenkins prediction for $M\simgt 10^{15}\Msun$. The {\it low-resolution} simulation MICE3072LR (solid line) and the {\it very-low resolution} simulation MICE3072CR (dashed line) evidence that our results are robust to mass resolution effects at large masses.}

\label{fig:massres}
\end{center}
\end{figure}

For a first glimpse of the abundance of massive objects in MICE, we display in Fig.~\ref{fig:massres} the mass function of FoF(0.164) halos obtained from our largest runs (in terms of simulated volume), including the corrections for mass and abundance discussed above in Secs.~(\ref{Sec:transients},\ref{sec:FoFmasscorrection}).

MICE3072 (blue squares) is in good agreement with the Jenkins prediction over more than two decades in mass, overlapping with the results from the larger MICE7680 (green circles) and MICE4500 (red triangles) for $M$ in $10^{14-15}\Msun$

However, as we transit towards the high-mass end ($M\simgt 10^{15} \Msun$) the abundance in the grand sampling volume of MICE7680 rises over the one in MICE3072 (and HVS) reaching a $20\%$ difference. 
In addition, measurements in MICE4500 (red triangles) are in very good agreement with those in MICE7680 even though these runs correspond to completely different initial conditions, softening length, box-size, etc (see Table~\ref{simtab}). 

We recall that MICE7680 and MICE4500 have roughly the same mass resolution to that in the HVS, but a volume $16.7$ and $3.4$ times larger, respectively. In turn, MICE3072 has roughly the box-size of the HVS, but $8$ times better mass resolution. 

To check that the ``excess'' abundance at large masses in not an artifact due to poor mass resolution we have included in Fig.~(\ref{fig:massres}) the mass function measured in MICE3072LR and in a {\it very-low} mass resolution run not listed in Table~\ref{simtab} ($L_{box}=3072\Mpc$, $N_p=512$, and $m_p = 1.5\times 10^{13}\Msun$). They both agree remarkably well with MICE3072  at $M\simgt 10^{15} \Msun$, showing that the abundance we found using MICE7680 and MICE4500 is robust to mass resolution effects, once the \pcite{warren06} correction is taken into account. 

In summary, we have tested the robustness of our results in front of several possible systematic effects. Early starting redshifts or inappropriate initial dynamics can affect the high-end mass function at the several $\%$ level. Hence the MICE simulations used either the 2LPT dynamics, or ZA with high-starting redshift ($z\sim 150$), with the exception of MICE3072 whose mass function we nonetheless correct as described in Sec.~\ref{Sec:transients}. In addition halo masses were computed using the correction of \pcite{warren06}, that we independently tested focusing in the regime of very massive halos, in order to avoid a bias towards higher masses at low particle number. Lastly, we studied whether the rather high particle mass of MICE7680 and MICE4500 can impact the abundance of massive halos by implementing a set of runs with fixed (large) volume and decreasing mass-resolution. The abundance of massive halos in these runs are in very good agreement once the \pcite{warren06} correction is considered, reinforcing the robustness of our mass function measurements in front of mass resolution effects.

\section{The abundance of FoF(0.2) halos}
\label{sec:micemf}

\begin{figure}
\begin{center}
\includegraphics[width=0.44\textwidth]{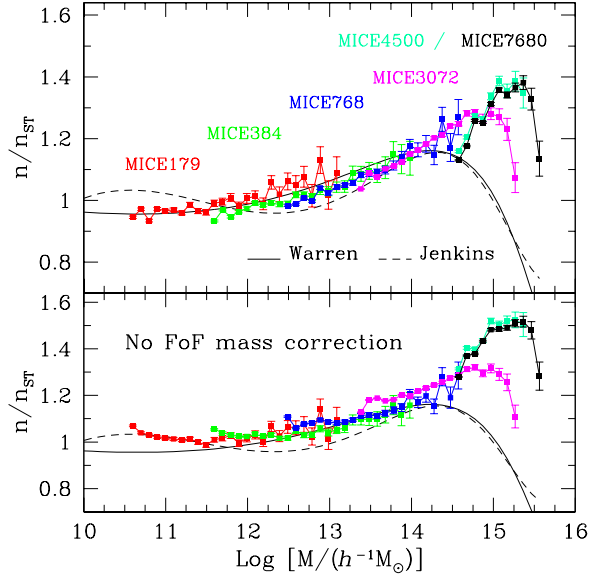}
\caption{The MICE {\it mass function at z=0,} measured after combining data from the set of MICE simulations with box-sizes $L_{box}=179$, $384$, $768$, $3072$, $4500$ and $7680 \Mpc$ (in red, green blue, magenta, sea-green and black respectively) and varying mass resolutions (see Table~\ref{simtab} for further details). Top panel corresponds to the measured mass function after applying the correction to the FoF mass as described in Warren et al. (2006). In the bottom panel we do not include this correction. In both figures we display the ratio to the Sheth and Tormen (1999) prediction and include the corresponding Warren and Jenkins fits for reference (solid and dashed lines). In each case the low-mass end is set by requiring a minimum of $100$ particles per halo while the high-mass by requiring a relative error below $5\%$ (displayed error bars correspond to Poisson shot-noise).}
\label{fig:MF_z0}
\end{center}
\end{figure}

Let us now turn to the mass function measurements in our catalogues of FoF(0.2) halos.
Figure~\ref{fig:MF_z0} shows the measured mass function in the MICE simulations tabulated in Table \ref{simtab}. We display the ratios to the Sheth \& Tormen fit in Eq.~(\ref{eq:ST}), binned in the same way as the measurements. Top panel corresponds to masses corrected for the FoF(0.2) bias as described in \pcite{warren06} and discussed in Sec.~\ref{sec:FoFmasscorrection}, Eq.~(\ref{eq:warrencorrection}). Bottom panel contains un-corrected mass functions. In both panels the solid line represents the Warren fit given in Eq.~(\ref{eq:warren}), while dashed corresponds to Jenkins fit in Eq.~(\ref{eq:jenkins}). The corrected mass functions agree very well with the Warren fit,  but only up to $10^{14}\Msun$. Past that mass there is a systematic underestimation of the halo abundance in MICE768 and MICE3072 that reaches $20\%$ at $M \sim 5\times 10^{14}\Msun$ (notice that we show only points with relative Poisson error $\le 5\%$). Part of this effect can be attributed to transients in the simulations used by \pcite{warren06} to calibrate the high-mass end, as discussed in \cite{2LPT,tinker08}.

For larger masses, $M \simgt M^{15} \Msun$, the underestimation of the Warren fit is even more severe, 
and grows monotonously with $M$. This in part might be due to volume effects:
the abundance of halos at the high-mass end is expected to be extremely low, 
of order $n_{halo}/\Mpccube \simlt 10^{-7}$ ($z=0$), $10^{-8}$ ($z=0.5$), and $10^{-9}$ ($z=1$) at $10^{15} \Msun$
(integrated over mass bins of $\Delta \log_{10} M=0.1$).
This means that already at moderate redshifts, $z=0.5$, a simulation of $L_{box} = 3 \Gpc$ will contain 
only about $300$ halos and therefore measuring the abundance of halos will be the subject to large uncertainties, i.e, 
the expected (shot-noise) error will be already of order $6\%$. 
By including larger volume simulations, such as MICE4500 ($L_{box}=4.5 \Gpc$) and MICE7680 ($L_{box}=7.68 \Gpc$), we
are able to increase the number of halos by up to a factor $\sim 16$, thus decreasing the associated 
halo abundance uncertainties by a factor of $\sim 4$.
As shown in Fig.\ref{fig:MF_z0} (top panel) results from both large volume simulations (MICE4500 \& MICE7680) agree
very well in the high-mass end ($M \simgt 3\times 10^{14}\Msun$). 
This agreement serves as a validation test for the implementation of each of them as well as 
for the high-mass end result, given that these two simulations share the same particle mass but have 
different initial dynamics (2LPT vs. ZA) and random phases.

\section{Error Estimation}
\label{sec:Errors}

In a rather general sense the most common source of statistical error considered in theoretical studies of halo abundance is solely the shot-noise contribution (e.g. \pcite{reed07,warren06,lukic07,HVS}). The importance of considering sample variance in addition to Poisson shot-noise had been highlighted in \cite{hu2002}, where it is shown that it can not be neglected in front of shot-noise for deriving precise cosmological constrains. Following this criteria, \pcite{tinker08} recently used jack-knife errors with the intention to account for both sampling variance at low mass and Poisson shot-noise at high ones.

To deepen into these considerations we will dedicate this section to perform a detailed study of different methods to estimate the error or variance in mass function measurements. One particular goal is to obtain well calibrated errors in order to implement an accurate fit  that could improve the high-mass description of Eq.~(\ref{eq:warren}). 

We will pay particular attention in comparing how internal errors (i.e. those derived using only the N-body for which the mean mass functions is measured, such as jack-knife) perform against external ones and theoretical predictions, depending on the mass regime and total simulated volume under consideration.

One of the {\it  internal} methods that we implemented is Jack-knife re-sampling \cite{zehavi2005}. For this we divided the simulation volume under consideration into $N_{JK}$ non-overlapping regions, and computed the halo number density in the full volume omitting one of these regions at a time. The variance (defined as the relative error squared) in the $i$-bin of the number density is then obtained as,
\beq
\sigma^{(i)^2}_{JK} = \frac{1}{\bar{n}^{(i)^2}}\frac{N_{JK}-1}{N_{JK}}\sum_{j=1}^{N_{JK}} (n^{(i)}_j-\bar{n}^{(i)})^2  
\eeq
where $\bar{n}^{(i)}$ is the mean number density of halos for that bin. In what follows we will show results using $N_{\rm JK}=5^3$, but we have checked that the estimates have already converged with varying $N_{\rm JK}$.

Another {\it internal} method we considered was to assume that the halos are randomly sampled and form a Poisson realization of the underlying number density field. In this case,
\beq
\sigma^{(i)^2}_{Poisson}=1/N_i
\eeq
where $N_i$ is the number of halos in the $i$ mass bin. 

For estimating the variance ${\it externally}$ in a volume $V$, we used an N-body of volume $V_L$, with $V_L >> V $. We then divide $V_L$ into several non-overlapping regions of volume $V$ and measure the number density in each sub-volume. This method, which we refer to as {\it sub-volumes}, is similar in spirit to boost-trap sampling except that the sub-volumes are not thrown at random and do not overlap. Thus, this method has the advantage of incorporating the effect of long-wavelength modes which are absent in the volume $V$. 

For example, for mass function errors in MICE179 we divided MICE384 in 8 sub-volumes and MICE768 in 80 sub-volumes. In this way, the best statistics for the error is achieved at the mid-to-high mass regime of MICE179 because the mass resolution of MICE768 does not allow to test all the way down to $M\sim 3.16\times 10^{11}\Msun$ (although MICE384 does). Nonetheless both MICE384 and MICE768 leads to a very consistent error estimation in MICE179 for its whole dynamic range, showing no dependence on mass resolution. For MICE384 we divided MICE768 in 8 sub-volumes and MICE3072 in 512 sub-volumes. The rest of box-sizes follow this same logic, that is, their variance in the mean number density was obtained from analyzing the next-in-volume runs as listed in Table \ref{simtab}.

Our last {\it external} method is ensemble average. This we can only apply to one box-size, $L_{box}=1200 \Mpc$, using the ensemble of $20$ independent realizations of MICE1200 as listed in Table \ref{simtab}.

Finally, to derive a theoretical estimate of the variance in the measured mass function  consider fluctuations in the mean number density of halos of a given mass, $\bar{n}_h(M)$, as coming from two different sources (see \pcite{hu2002} for the original derivation). Firstly, a term arising from fluctuations in the underlying mass density field $\delta_m$, if we consider the halo number density to be a tracer of the mass. If this relation is simply linear and local then $\delta n_h (M,{\bf x})/ \bar{n}_h = b(M) \delta_m({\bf x})$, where $b$ is the halo bias. 

Secondly, a shot-noise contribution $\delta n_{\rm sn}$ due to the imperfectness of sampling these fluctuations with a finite number of objects. This noise satisfies $\langle \delta n_{\rm sn} \rangle = 0 $ and is assumed to be un-correlated with $\delta_m$. Furthermore,  if we assume the halo sample to be a Poisson realization of the true number density this error becomes a simple Poisson white-noise with variance $\langle \delta n^2_{\rm sn} \rangle / {\bar n}^2_h = 1/ {\bar n}_h V= 1/N$, where $N$ is the total number of objects sampled within the volume $V$. Within these assumptions we then have,
\beq
\delta n_h  (M,{\bf x}) = b(M) \bar{n}_h(M) \delta_m ({\bf x}) + \delta n_{\rm sn}.
\label{eq:deltan}
\eeq
The number density of objects of this mass within the simulation is estimated by,
\beq
n = \int_V d^3 x W({\bf x}) \left[ \bar{n}_h + \delta n_h \right],
\label{eq:nint}
\eeq
where $W({\bf x})$ is the simulation window function, normalized such that $\int_V d^3 x W(x)=1$. The variance of the number density measurements in the simulation is then given by,
\begin{eqnarray}
\langle n^2 \rangle-\bar{n}_h^2&=&\frac{\bar{n}_h}{V}+b^2\bar{n}_h^2 \int d^3 x_i \int d^3 x_j W({\bf x}_i) W({\bf x}_j) \nonumber \\
&&\times\langle \delta_m({\bf x}_i) \delta_m({\bf x}_j)\rangle,
\label{eq:variance}
\end{eqnarray}
and can be cast as,
\beq
\sigma_h^2=\frac{\langle n^2 \rangle - \bar{n}_h^2}{\bar{n}_h^2}=\frac{1}{{\bar n}_h V} + b^2_h \int \frac{d^3k}{(2\pi)^3} |W(k R)|^2 P(k),
\label{eq:sigmaMF}
\eeq
where $P(k)$ is the linear power spectrum of mass. For simplicity, we will assume the simulation window function $W$ to be top-hat in real space, Eq.~(\ref{eq:window}), with smoothing radius such that the window volume equals the simulated one, i. e. $R=(3V/4\pi)^{1/3}$.

The first term in Eq.~(\ref{eq:variance}) is the usual shot-noise contribution to the variance, that we introduced rather {\it had-hoc} in Eq.~(\ref{eq:deltan}), but it can also be derived in the context of the halo model as the contribution from the 1-halo term \cite{takada2007}. The second term, know as sampling variance, is the error introduced by trying to estimate the true number density using a finite volume.

As discussed in Sec.~\ref{sec:binmf} one is in practice interested in the halo abundance within bins of mass range $[M_1-M_2]$ and characteristic mass $M$. Thus in Eq.~(\ref{eq:sigmaMF}) we will compute $M$ and $\bar{n}_h$ from Eqs.~(\ref{eq:n},\ref{eq:m}) and the bias as,
\beq
b_h=\frac{1}{{\bar n}_h} \int_{M_1}^{M_2} b_{ST}(M) (dn/dm) \, dM,
\label{eq:bias}
\eeq
where $b_{ST}$ is the prediction for the linear bias dependence on halo mass from \pcite{ST99},
\beq
b_{ST}(M)= 1 + \frac{q \delta_c^2/\sigma^2-1}{\delta_c}+\frac{2p/\delta_c}{1+(q \delta_c^2/\sigma^2)^p},
\label{eq:bST}
\eeq
with parameters $q=0.707$ and $p=0.3$ and $\sigma=\sigma(M)$ given in Eq.~(\ref{eq:sigma8}). Equation~(\ref{eq:bST}) follows from considering variations of the unconditional mass function in Eq.~(\ref{eq:ST}) with respect to the critical over-density for collapse $\delta_c$. Thus, strictly speaking this bias expression should be weighted by $dn/dm$ from Eq.~(\ref{eq:ST}) when integrating Eq.~(\ref{eq:bias}) \cite{ST99,manera09}. However we have found that using a fit to the MICE mass function instead leads to better agreement with measurements of clustering in the simulations. Accordingly we will use this fit also to estimate $M$ and $\bar{n}_h$, entering in Eq.~(\ref{eq:sigmaMF}). 

\begin{figure}
\begin{center}
\includegraphics[width=0.5\textwidth]{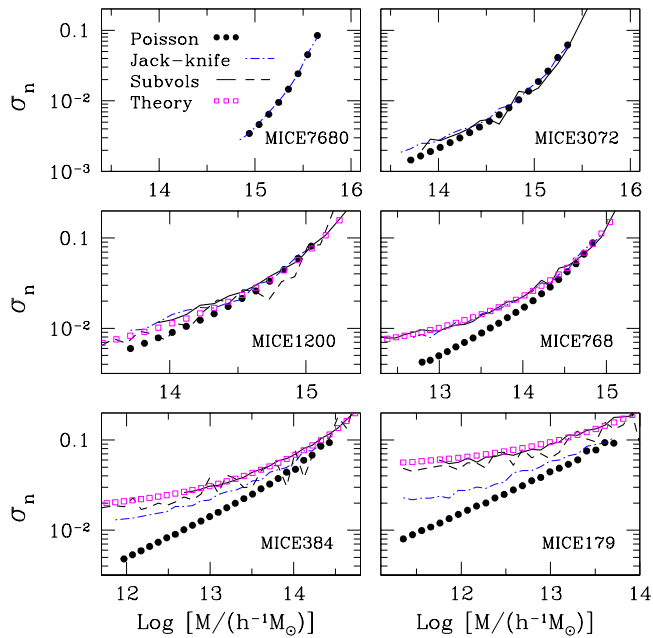}
\caption{{\it Sampling variance vs. shot-noise:} The panels show different estimates for the variance in the halo mass function within each of the $6$ box-sizes used throughout this paper. Dot symbols correspond to Poisson shot-noise and is only shown for halos with a minimum of $200$ particles and until the relative error reaches $10\%$ (as used in Sec.~\ref{sec:fit}). Dot-dashed blue line is the result of implementing jack-knife re-sampling in the given box-size.
Solid and dashed black lines are the sub-volumes method described in the text that uses independent sub-divisions of larger-size boxes. 
For this method we include cases in which two larger volume runs are used (panels of MICE179, MICE384 and MICE1200). Empty magenta squares are the theoretical estimate in Eq.~(\ref{eq:sigmaMF}) that includes  both sampling and shot-noise variance.
The variance is shot-noise dominated roughly for $M>10^{14}\Msun$ (depending on the box-size). While jack-knife does capture some sampling variance at smaller masses, it is not fully satisfactory. It can under-estimated the error by a factor of $2-3$ at $10^{12} \Msun$. Notably, the sub-volumes and theoretical estimates are in very good agreement across all box-sizes and full mass range.}

\label{fig:MF_ErrorsJK}
\end{center}
\end{figure}

Figure~\ref{fig:MF_ErrorsJK} shows the result of our {\it external}, {\it internal} and {\it theoretical} error study. Clearly the Poisson shot-noise dominates the error budget for $M \simgt 10^{14}\Msun$ corresponding to $\sigma_{MF}/MF > 5\%$ (and $\log \sigma^{-1} \simlt 0.06$). At smaller masses the sampling variance becomes increasingly important rapidly dominating the total error (this is more so for smaller box-sizes). Jack-knife re-sampling does capture this trend but only partially, in particular for the smaller box-sizes ($\simlt 500 \Mpc$) where sampling variance from the absence of long-wavelength modes is more significant. This seems to indicate that the jack-knife regions must have a minimum volume (e.g. while jack-knife works well at $10^{13}\Msun$ in MICE768, it does not for MICE179 at the same mass). This is an important result to consider in further studies where jack-knife re-sampling is used to improve upon Poisson shot-noise. The total error can be under-estimated by a factor of a few, e.g. $3$ ($1.5$) at $10^{12}\Msun$ ($10^{13}$) in MICE179. 

On the other hand, the theoretical error from Eq.~(\ref{eq:sigmaMF}) is in remarkably good agreement with the sub-volumes method described above across all box-sizes (not shown in MICE7680 and MICE3072 panels of Fig.~\ref{fig:MF_ErrorsJK} because is indistinguishable from the other estimates). This can be taken as a cross-validation of these two methods. 

In addition we have tested how these different methods compare with the ensemble error in the mass function obtained from the $20$ independent runs of MICE1200. For the sub-volumes estimation we divided MICE7680 in $264$ regions of volume almost identical to that of MICE1200 (as well as MICE3072 in $18$ regions for consistency checks). The result is that, for $L_{box}=1200\Mpc$, the sub-volume error is larger than the ensemble one by a factor of about $20\%$ at $M\sim [10^{13}-3.16\times 10^{14}] \Msun$. The reason is that each sub-volume region ``suffers'' fluctuations in the mean density caused by the long-wavelength modes present in the larger box-size from which they have been obtained (MICE7680 or MICE3072 in this case). These modes, that introduce an extra-variance, are absent in each of the ensemble members that satisfy periodic boundary conditions at the scale $L_{box}=1200\Mpc$. Thus, the ensemble error (from running different simulations) does not always work well because it  can suffer from volume effects.

This conclusion can be nicely reinforced using  Eq.~(\ref{eq:sigmaMF}), that performs very well here also.
One can mimic the absence of long-wavelength modes by setting the low-$k$ limit in the sampling variance integral in Eq(\ref{eq:sigmaMF}) to be the fundamental mode of MICE1200 ($k_f = 2\pi/ 1200 \kvecMpc$). In this case the theoretical model agrees with the ensemble error (in fact, $\sigma_h$turns to be mostly dominated by shot-noise). Instead, by setting it to $k_f\sim0$ one recovers the sub-volumes estimate. 

In summary, the sub-volume method should be considered as the one comprising all statistical uncertainties: shot-noise, sampling variance and volume effects (the fact that there are fluctuations in scales larger than the sample size).
The theoretical estimate in Eq.~(\ref{eq:sigmaMF}) is consistent with it to a remarkably good level, for the masses and box-sizes tested in this paper.
Hence, is a powerful tool for studies involving the abundance of massive halos \cite{rozo09,vikhlinin09}.

\section{The Fitting function for the abundance of massive halos}
\label{sec:fit}

The accurate sampling of the mass function requires a demanding combination of very big volumes 
and good mass resolution. As shown in Fig.\ref{fig:micesims}, using the MICE set of simulations 
we have sampled volumes up to $450 \Gpccube$ with a wide range of mass resolutions
yielding a dynamic range $10^{8} < M/(\Msun) < 10^{12}$ in particle mass (see also Table~\ref{simtab}). 

As shown in Fig.~\ref{fig:MF_z0}, the ST and Warren fits underpredict the abundance 
of the most massive halos found in our N-body simulations for $M > 10^{14}\Msun$, 
although Warren gives accurate results for lower masses. 

In this section we derive new fits based on the MICE set of simulations,
sampling the mass function over more than 5 orders of magnitude in mass, and covering
the redshift evolution up to $z=1$. 
For this purpose we use a set of MICE simulations with increasing volume and corresponding decreasing mass resolution,
in order to sample the mass range from the power-law behavior at low masses, $M\sim10^{10}\Msun$ and up to the exponential cut-off at the high-mass end.
As discussed in Sec.~\ref{sec:micemf}, the abundance of halos at the high-mass end is expected to be extremely low, 
of order $n_{halo}/\Mpccube \simlt 10^{-7}$ ($z=0$), $10^{-8}$ ($z=0.5$), and $10^{-9}$ ($z=1$) at $10^{15} \Msun$
(integrated over mass bins of $\Delta \log_{10} M=0.1$), and thus we shall use MICE4500 ($L_{box}=4.5 \Gpc$) 
and MICE7680 ($L_{box}=7.68 \Gpc$) in order to get a more accurate measurement of the halo abundance in this regime.
In practice, we shall combine cluster counts from both simulations to get a more robust estimate.
As it will be shown below, our fitting functions recover
the measured mass function over the entire dynamic range and its redshift evolution with
$\sim 2\%$ accuracy.

\begin{figure}
\begin{center}
\includegraphics[width=0.44\textwidth]{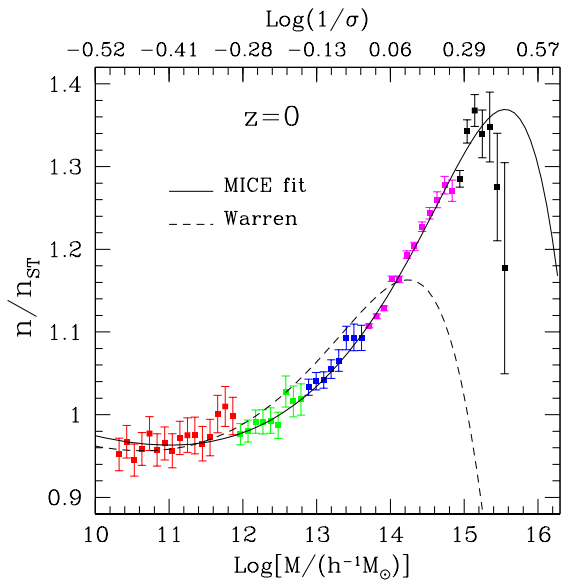}
\caption{{\it Mass function fit at z=0.} 
Symbols as in Fig.~\ref{fig:MF_z0}.
Measurements sample the mass function over more than 5 orders of magnitude,  
$4\times 10^{15}> M/\Msun > 2\times 10^{10}$.
Results are ratioed to the ST fit. 
The best-fit to the N-body measurements (``MICE'' fit, solid line) 
is given by the Warren-like mass function, Eq~(\ref{eq:warren}),
with parameters as given in Table~\ref{fittab}. 
The fit agrees with N-body data to 2$\%$ accuracy
in practically all the dynamic range ($2.5 \times 10^{15} > M/\Msun > 2\times 10^{10}$).
The Warren fit (dashed line)  matches the N-body to 3$\%$ accuracy in the low mass end
$M/\Msun < 10^{14}$, but it significantly underestimates the abundance of the most massive halos:
we find a $10\%$ ($25\%$) underestimate at $M/\Msun \sim 3\times 10^{14} ~(10^{15})$, and larger biases
for more massive objects.}  
\label{fig:MFfit_z0}
\end{center}
\end{figure}

For our fitting procedure we use the following simulations: 
MICE179,384, 768, 3072, 4500 and MICE7680 (see Fig.\ref{fig:micesims} and Table\ref{simtab})
and match them so that in overlapping mass bins we shall adopt the abundance estimated from the simulation with the 
lower associated error, provided halos include a minimum of 200 particles, except for the smallest box-size simulation, MICE179, for which we use down to 50 particle halos. 
This is done in order to 
sample small enough halos (as small as $10^{10} \Msun$) whose abundance has been accurately
measured in previous analyses (see e.g, \pcite{warren06,tinker08}) 
and that we aim at recovering as well with our fitting functions.

On the other hand, for $M \simgt 3\times 10^{14}$ we 
average results from both MICE4500 and MICE7680. As shown above, for $z=0$, 
results for $M > 10^{15}\Msun$ from MICE7680 are found to be in 
full agreement with those of MICE4500 
despite the different initial conditions and time-step size used, 
what provides a robustness test to our measurements at the highest mass bins.

The fit to the mass function is then determined using a diagonal $\chi^2$ analysis,
\begin{equation}
\chi^2=\sum_{i=1}^{N} \frac{{(n^{(i)}_{fit}-n^{(i)}_{Nbody})}^2}{{\sigma^{(i)}}^2}
\label{chi2}
\end{equation}
where $n^{(i)}_{fit}$ ($n^{(i)}_{Nbody}$) is the theoretical (N-body) mass function integrated over 
the i-th logarithmic mass bin of width $\Delta \log_{10} M /(\Msun)= 0.1$, and the errors $\sigma^{(i)}$ 
are computed using the Jack-knife (JK) estimator. 
The JK estimator is consistent with theoretical errors and 
sub-volumes dispersion (see Fig.~\ref{fig:MF_ErrorsJK}), 
except for the smallest mass bins 
(sampled by MICE179, see lower right panel in Fig.~\ref{fig:MF_ErrorsJK}) 
for which the JK errors significantly underestimate other error estimates.
By using JK errors for those bins we just give them a larger statistical weight in the $\chi^2$ analysis, thus making sure the fit recovers the expected low mass behavior (i.e we assume something similar to a {\it low-mass prior}).

We summarize the fitting results for the MICE mass functions 
at $z=0$ and $0.5$ in Table \ref{fittab}.
The best-fit to the N-body measurements at $z=0$ (``MICE'' fit, solid line) 
is given by the Warren-like mass function, Eq.~(\ref{eq:warren}),
with parameters $A=0.58$, $a=1.37$, $b=0.30$, $c=1.036$ with $\chi^2/\nu=1.25$.
As shown in Fig.~\ref{fig:MFfit_z0}, the fit recovers our N-body data to 2$\%$ accuracy
in practically all the dynamic range, that is, for $2.5\times 10^{15} > M/(\Msun) > 4\times 10^{10}$. 
The Warren fit (dashed line)  matches the N-body to the same accuracy 
in the mass regime $M < 10^{14} \Msun$, 
but significantly underestimates the abundance of the most massive halos:
we find a $10\%$ ($30\%$) underestimate at $M =  3.16 \times 10^{14}\Msun$ ($10^{15}\Msun$).
This can be attributed in part to transients in the \cite{warren06} N-body data, as discussed in Sec.~\ref{sec:micemf}.

\begin{figure*}
\begin{center}
\includegraphics[width=0.45\textwidth]{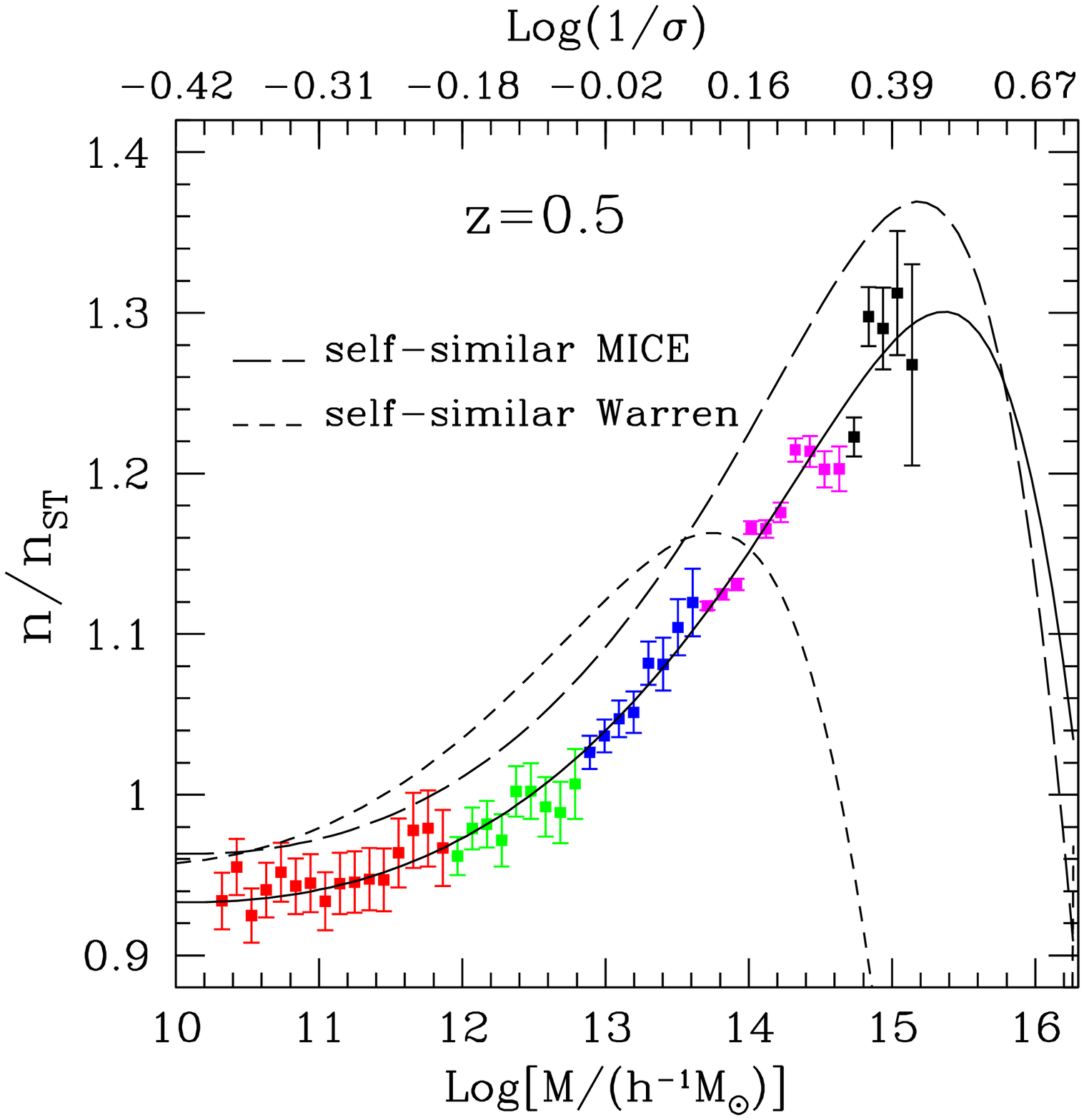}
\includegraphics[width=0.45\textwidth]{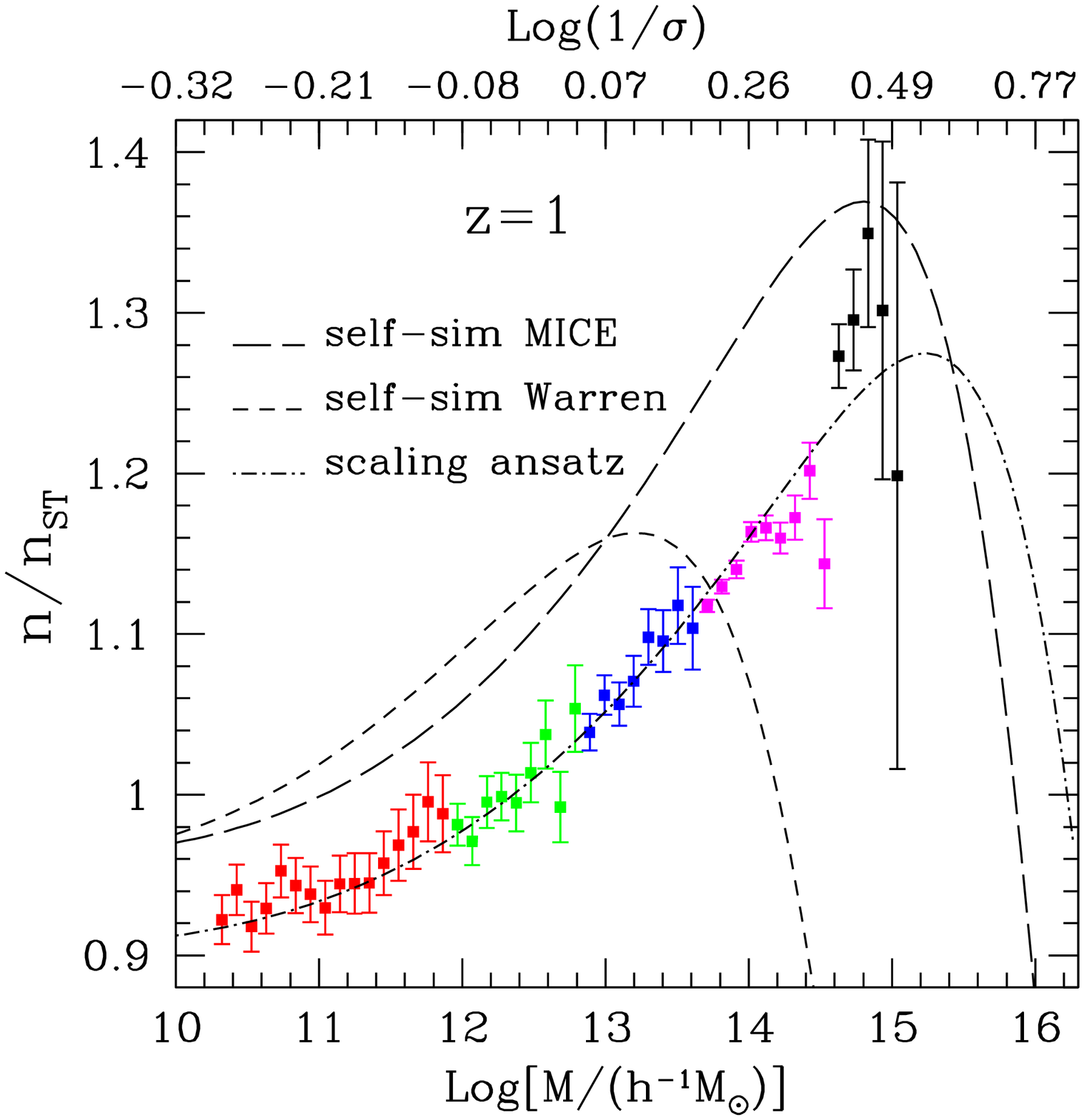}
\caption{{\it Left: Mass function fit at z=0.5.} 
The best-fit to the N-body measurements is given by Eq.~(\ref{eq:warren})
with parameters as given in Table~\ref{fittab} (solid line).
This fit matches simulations to 2$\%$ accuracy
in all the dynamic range ($3\times 10^{14} > M/\Msun > 2\times 10^{10}$).
Assuming a {\it self-similar} form of the mass function 
(i.e, using the N-Body fit at $z=0$ extrapolated with the linear growth to $z=0.5$), 
one overestimates the measurements by $3-6\%$ as the halo mass increases (see dashed line).
The (self-similar) Warren fit yields an even larger overestimate of the halo abundance ($3-10\%$).
This shows to what extent the FoF(0.2) mass function deviates from universality.
{\it Right: Mass function fit at z=1.} 
Growth of the mass function with redshift can be accurately modeled with a simple ansatz,
Eqs.(\ref{eq:mfzans}),(\ref{eq:alpha}),(\ref{eq:micefit}) (see discussion below, in Sec.\ref{sec:MF_growth}), 
and it provides a good fit to simulations.
We find that deviations from universality of the halo abundance increase with redshift: 
the self-similar MICE fit exceed by up to $10\%$ the simulation measurements (and similarly for the Warren fit).}
\label{fig:MFfit_highz}
\end{center}
\end{figure*}

At higher-redshifts, the mass function deviates from the
universal form in Eq.~(\ref{eq:warren}), and the fitting parameters change accordingly.
In particular, for $z=0.5$ we find the best-fit values
$A=0.55$, $a=1.29$, $b=0.29$, $c=1.026$, yielding a $\chi^2/\nu=1.20$. 
As seen in Fig.~\ref{fig:MFfit_highz} (left panel), this fit recovers the N-body measurements 
to 2$\%$ accuracy in all the dynamic range (i.e., for $10^{15} > M/(\Msun) > 2\times 10^{10}$).
The MICE fit at $z=0$ extrapolated with the linear growth to $z=0.5$ ($z=1$) 
overestimates the measurements by $3-6\%$ ($10\%$) as the halo mass increases. This in turn shows
to what extent the FoF(0.2) mass function deviates from universality.

\begin{table}
\begin{center}
\begin{tabular}{lcccccccccc}
\hline \\
$z$  &&   $A$ &&   $a$ &&   $b$ && $c$  && ${\chi^2/\nu}$ \\
\hline \\
$0$   &&    $0.58$  && $1.37$      && $0.30$  &&  $1.036$  && $1.25$     \\ 
$0.5$   &&    $0.55$  && $1.29$    && $0.29$   &&  $1.026$ && $1.20$   \\
\hline
\end{tabular}
\end{center}
\caption{Mass function best-fit parameters for $f(\sigma,z)$, Eq.~(\ref{eq:warren}), and goodness of fit.}
\label{fittab}
\end{table}


\section{Halo growth function}
\label{sec:MF_growth}

In the previous section we found that the mass function deviates significantly from universality (or self-similarity).
Here we investigate in detail the {\it halo growth function}, i.e,  the evolution of the halo abundance with redshift,
using the scaling evolution of the best-fit parameters as a starting point. 

The evolution of the fitting function parameters with redshift, as shown in Fig.~\ref{fig:MFfit_highz}, 
indicate that the FoF(0.2) mass function, at least for the linking length $l=0.2$, is non-universal at the $5\%-10\%$
level in the red-shift range tested.  
This is in agreement with previous studies \cite{HVS,reed03,reed07,lukic07,tinker08} but is 
consistently extended here to the high-mass regime, 
hard to sample robustly particularly at higher red-shifts. 

To account for this evolution we next try to fit the mass function growth with a simple ansatz.
If we follow \cite{tinker08} and assume that the fitting parameters are a simple function 
of the scale factor, $a=1/(1+z)$, we can model the evolution as,
\begin{equation}
P(z) = P(0)(1+z)^{-\alpha_i} \ ; P=\{A,a,b,c\} \ ; \alpha_i=\{\alpha_1,\cdots,\alpha_4\}
\label{eq:mfzans}
\end{equation}
where $P(0)$ are the fitting parameters at $z=0$, as given by Table~\ref{fittab}.
Therefore, we can use the lowest redshift measurements at $z=0$ and $z=0.5$ to determine
the slope parameters $\alpha_i$,
\begin{equation}
\alpha_1=0.13 , \ \alpha_2=0.15, \ \alpha_3=0.084, \ \alpha_4=0.024.
\label{eq:alpha}
\end{equation}
If the ansatz is correct, i.e, the growth of the mass function 
can be modeled to a good approximation with Eq.~(\ref{eq:mfzans}), one should be able to {\it predict}
the measured cluster evolution at higher redshifts. 
Using the values of $\alpha_i$ as given above, we {\it predict} the following fitting parameters
at $z=1$: $A=0.53$, $a=1.24$, $b=0.28$, $c=1.019$, what gives a good match to simulations, $\chi^2/\nu=1.92$.
As shown in both panels of Fig.~\ref{fig:MFfit_highz}, the scaling  
ansatz recovers the measured mass function to 3$\%$ accuracy
in most of the dynamic range (i.e, for $3.16\times 10^{14} > M/(\Msun) > 2\times 10^{10}$).

We conclude from this that the ansatz  Eq.~(\ref{eq:mfzans}) can be safely used to make predictions about 
the abundance of the most massive halos at intermediate redshifts.

It has been argued by \cite{tinker08} that the non-universality of the mass function is basically a 
consequence of the evolution of the halo concentrations, which in turn is mostly due to the change 
of the matter density $\Omega_m$ with redshift and thus $f(\sigma)$ 
should be rather modeled as a function of the linear growth rate of density perturbations, $D(z)$. 
In order to test this hypothesis, we have repeated the analysis where the scaling of $f(\sigma)$   
is parametrized as follows,
\begin{equation}
P(z) = P(0)(D(z)/D(0))^{\beta_i} \ ; \ \beta_i=\{\beta_1,\cdots,\beta_4\}
\label{eq:mfzans2}
\end{equation}
In this case the slope parameters are found to be,
$\beta_1=0.22 , \beta_2=0.25, \beta_3=0.14, \ \beta_4=0.04$.
Using this model we estimate $f(\sigma)$ parameters at $z=1$ to be, $A=0.52$, $a=1.22$, $b=0.28$, $c=1.017$, 
what provides a slightly worse fit to the the N-body measurements, with $\chi^2/\nu = 2.85$.
Therefore, we find some evidence in favor of a scaling ansatz based on the scale factor with respect to 
that based on the growth rate. We note that our analysis is of limited validity since we have only 
considered one cosmology and one should explore a wider parameter space to draw stronger conclusions
on this point. 

In summary, the fit to the FoF(0.2) halo mass function measured in the MICE simulations between redshift $0$ and $1$ 
is given by,
\beq
f_{\rm MICE}(\sigma,z)= A(z) \left[\sigma^{-a(z)} +b(z)\right] \exp\left[-\frac{c(z)}{\sigma^2}\right]
\label{eq:micefit}
\eeq
with $A(z)=0.58(1+z)^{-0.13}$, $a(z)=1.37(1+z)^{-0.15}$, $b(z)=0.3(1+z)^{-0.084}$, $c(z)=1.036(1+z)^{-0.024}$. 

\begin{figure}
\begin{center}
\includegraphics[width=0.48\textwidth]{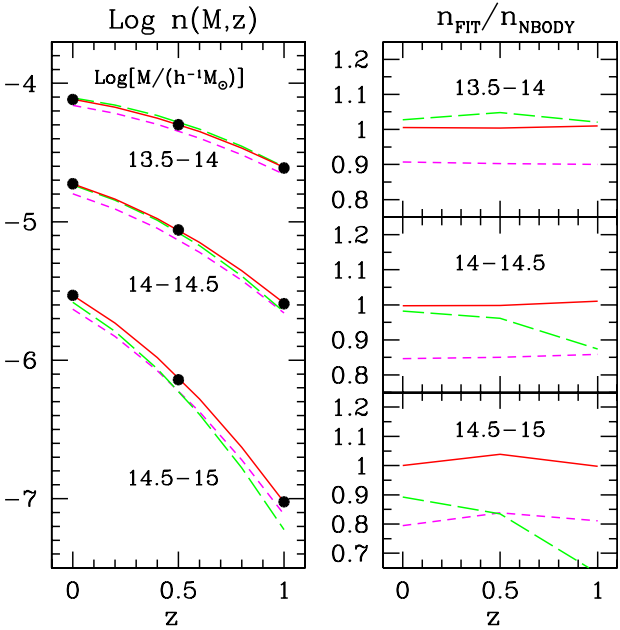}
\caption{{\it Halo Growth Function:} 
evolution of the halo abundance with redshift. We show the three most massive logarithmic 
bins (with $\Delta logM/(\Msun) = 0.5$), where we find the most significant deviations 
between N-body and available fits.
{\it Left panel:} the abundance of massive halos drops by half (one) order of magnitude 
for the $log M/(\Msun) = 13.5-14 ~(14.5-15)$ from z=0 to z=1, in rough agreement with 
analytic fits.
{\it Right panel:}. Residuals between analytic fits (lines) to N-body measurements (symbols)
for the same mass bins (increasing mass from top to bottom panels).
ST and Warren fitting functions (short and long-dashed lines respectively) only match 
the high-mass end of the measured mass function at z=0.5 to $15\%$ accuracy. The extrapolated Warren fit to z=1 underestimates MICE data by up to $30\%$. The {\it predicted} halo abundance growth 
from the MICE fits at low redshift (red solid line; see text for details) 
recovers the measured abundance at z=1 to better than $1\%$.}
\label{fig:MFz}
\end{center}
\end{figure}

We can now explore how the {\it halo growth function} evolves in more detail.
For this purpose, we study the halo mass function integrated in wider logarithmic bins 
($\Delta \log_{10} M/(\Msun) = 0.5$), and concentrate on the highest mass bins where we find the largest 
deviations between our N-body measurements and available fits. Figure~\ref{fig:MFz} (left panel) 
shows the halo growth factor as measured in several comoving redshifts in the MICE simulations.

Our simulations show that the abundance of massive halos drops by half (one) order of magnitude 
for the $logM/(\Msun) = 13.5-14 ~(14.5-15)$ from z=0 to z=1, 
in rough agreement with analytic fits.
However, as displayed in the right panels of Fig.~\ref{fig:MFz}, 
the (self-similar) ST and Warren fitting functions (see short and long-dashed lines respectively) only match 
the high-mass end of the measured mass function at z=0.5 to $15\%$ accuracy. 
The Warren fit at z=1 underestimates simulation data by up to $30\%$. 
On the other hand, using the {\it predicted} halo abundance growth from the MICE fits at low redshift 
(red solid line) recovers the measured abundance at z=1 to better than $1\%$. We have used the 
scaling functions given by Eq.~(\ref{eq:mfzans}), however we have checked that our results do not
change significantly if we use the growth rate ansatz instead, Eq.~(\ref{eq:mfzans2}).

\section{Cosmological Implications: Bias on Dark-energy Constraints}
\label{sec:cosmo}

The cluster mass function is one of the standard cosmological probes used
by current and proposed surveys to constrain cosmological parameters.
In particular, the cluster abundance as a function of cosmic time 
is a powerful probe to determine the nature of dark-energy.
However usage of the cluster abundance as a cosmological probe is limited by
systematics in the mass-observable relations and the potential impact of priors (see e.g, \pcite{battye03,weller03}).

Here we concentrate on the impact of priors in the mass function on extracting the dark-energy equation
of state, $w$. As shown in \ref{sec:MF_growth}, the halo mass function measured in simulations deviates form 
self-similarity by as much as $15\%$, depending on redshift and halo mass.
We shall estimate how this systematic departure from universality can potentially {\it bias} estimates of $w$. 

As a working case, we will consider a tomographic survey with photometric accuracy $\Delta z$, and
estimate the shift in the recovered value of, $w$, 
by including cluster counts in redshift shells up to a given 
depth, $z$. For simplicity we consider a constant dark-energy equation of state, although this same
analysis could be easily extended to a time varying $w(z)$.
Since the {\it wrong prior} on the halo mass function 
can be mistaken by the {\it right} cluster abundance for a {\it different} cosmology, 
the bias on $w$, for a given redshift, will be determined by the relative sensitivity on $w$ 
of the two cluster count probes of dark-energy: the mass function growth and the survey volume up to a given depth. 

We perform a $\chi^2$ analysis to determine the bias as a function of survey depth 
by comparing halo counts in redshift shells as follows,
\beq
\chi^2=\sum_{z_i} \frac{(n(w)^{(i)}-n(z)^{(i)}_{Nbody})^2}{{\sigma^{(i)}}^2},
\label{biaschi2}
\eeq
where $n(w)^{(i)}$ are the counts from the assumed {\it self-similar} mass function, 
for a cosmology with a given value of $w$, integrated from a minimum mass $M_{min}$ 
up to some maximum mass $M_{max}$, for the redshift shell $z_{i}$ of width $\Delta z_{i}$ within which we 
can safely consider the mass function to be independent of $z$. 
In turn, $n(z)_{Nbody}$ are the 
corresponding counts measured in simulations that use the fiducial cosmology with $w=-1$, 
and that are accurately described by the scaling ansatz, Eq.~(\ref{eq:micefit}).
The associated error is assumed to be pure shot-noise given by the N-Body counts, $\sigma = \sqrt{n_{Nbody}}$.
Here we make estimates for full-sky surveys (i.e, we will draw optimistic forecasts for a given survey depth), 
although smaller areas can be easily incorporated in our analyses by scaling the shot-noise error accordingly.
We shall consider surveys with SZ detected clusters, i.e, with a redshift independent 
mass threshold $M \sim 10^{14} \Msun$, and constant photo-z error $\Delta z=0.1$. 
The upper limit in the mass is taken to be $10^{15.5}\Msun$ to avoid
possible systematic departures of the N-Body fit used beyond this mass-cut with respect 
to the simulation measurements. However we have checked our results do not change if we take larger mass cuts.

\begin{figure}
\begin{center}
\includegraphics[width=0.5\textwidth]{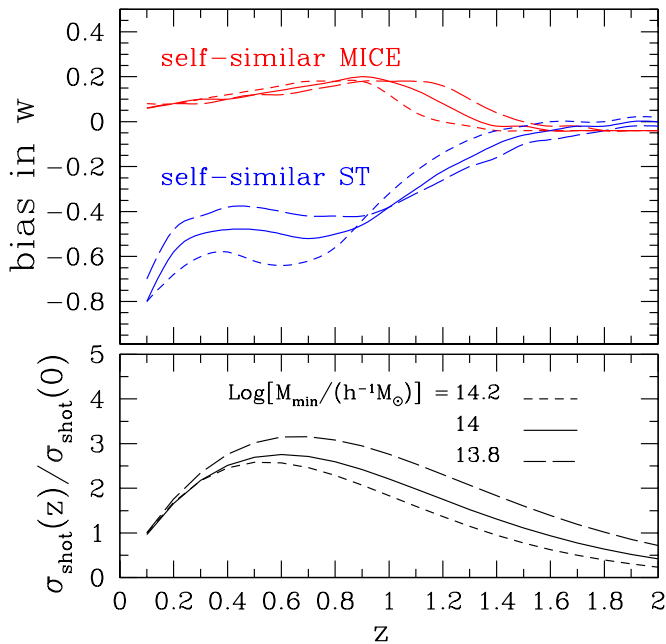}
\caption{{\it Bias on $w$ induced by self-similar prior on the mass function}.
{\it Upper panel:} 
Assuming the self-similar ST (MICE) fit induces up to $50\%$ ($20\%$) shift in the recovered value of $w$
for shallow or medium depth surveys $z \simlt 1$, whereas the effect tends to be negligible for deep surveys.
{\it Bottom panel:} the relative shot-noise error decreases with redshift, what makes high-z cluster counts, 
which are dominated by the strong variation of the mass function with $w$, 
to down-weight biases induced by the low-z counts.}
\label{fig:wbias}
\end{center}
\end{figure}

Figure~\ref{fig:wbias} shows the bias on $w$ as a function of survey depth $z$ for 
two different priors on self-similar mass functions, the ST fit and the MICE fit (i.e, assuming the
fit at $z=0$ extrapolated to higher-z). 
We find that the estimated bias is robust to better than $20\%$ to changes in the SZ mass threshold $M_{min}$ 
from $10^{13.8}$ to $10^{14.2}$ (see dashed lines in Fig.\ref{fig:wbias}).  
For the ST fit, the bias can be as large as $50\%$ for survey depth $z \simlt 1$, whereas for the MICE 
self-similar fit it reduces to $20\%$ at most for the same depth. The bias at low-z results from the 
relatively small and comparable sensitivities to changes in $w$ of the geometric (volume) and the shape of the 
mass function growth.
On the other hand, the strong bias at low-z for the ST prior is due to the poor 
fit it gives to N-Body for the relevant masses $M \simgt 10^{14}\Msun$.
This systematic effect tends to decrease for increasing depth as the mass function growth becomes a much stronger 
function of the dark-energy equation of state than the redshift shell volume (for constant $\Delta z$)
and thus it determines the cluster counts irrespective of the prior on the mass function. 
This trend is strengthen by the fact that the shot-noise error per z-shell drops with depth as well (see bottom panel of Fig.\ref{fig:wbias}), so the
high-z counts down-weight the observed low-z bias on $w$. 
Therefore, for deep surveys $z \simgt 1$ any bias associated to the mass function
tends to be washed out.

\section{Discussion and Conclusions}
\label{sec:Conclusions}

The abundance of clusters as a function of cosmic time is a powerful 
probe of the growth of large scale structure. In particular, clusters 
provide us with a test of the sensitivity of the growth of structure 
to cosmological parameters such as the dark-matter and dark-energy density content,
its equation of state, or the amplitude of matter density fluctuations. 
More importantly, the abundance and clustering of clusters are 
complementary to other large scale structure 
probes such as the clustering of galaxies, weak lensing and supernovae 
(see e.g, \pcite{hu2002,cunha2009,cunhaetal2009,estrada09}).

In order to plan and optimally exploit the scientific return of upcoming astronomical surveys
(e.g, DES, PAU, PanSTARRS, BOSS, WiggleZ, Euclid) one needs to make accurate forecasts of the
sensitivity of these probes to cosmological parameters in the presence of systematic effects.
Accurate determination of the abundance of clusters is limited by our knowledge 
of the relation between cluster mass and observable (see e.g,~\pcite{hu2003}). One key ingredient
in this mass-observable calibration is the precise determination of the halo mass function.

In this paper we use a new set of large simulations, including up to $2048^3 \simeq 10^{10}$ 
particles, called MICE (see Table \ref{simtab}) to accurately determine the abundance of massive halos 
over 5 orders of magnitude in mass, $M = (10^{10}-10^{15.5}) \Msun$,  
and describe its evolution with redshift.
Notice that the MICE simulations sample cosmological volumes
up to $7.68^3 \simeq 450 \Gpccube$. Armed with these simulations 
we estimate, for the first time, 
the mass function over a volume more than one order of magnitude 
beyond the Hubble Volume Simulation size, $3^3 = 27\Gpccube$ \cite{HVS,HVS2}. 
Our results have been tested against mass resolution, choice of initial conditions,
and simulation global time-step size.

Our findings can be summarized as follows:
\begin{itemize}

\item{We confirm previous studies (see \pcite{2LPT}) showing
that an accurate determination of the mass function is sensitive
to the way initial conditions are laid out. For a fixed starting
redshift, the usage of 2LPT instead of the ZA to set-up the initial 
conditions helps to avoid the effects of transients that
tend to artificially decrease the abundance of large-mass halos. Alternatively,
one can use ZA with a high enough starting redshift. 
In particular, we find an under-estimate of the halo abundance by $\sim 5\%$ at $M \sim 10^{15}$, $3.1\times 10^{14}$ and $10^{14}\Msun$  at $z=0$, $0.5$ and $1$ respectively 
(and larger for larger masses, at fixed redshift) if ZA $z_i=50$ is used instead of 2LPT.
All our mass functions were safe from transients or corrected accordingly for those simulations that need it
(i.e MICE3072, see Table~\ref{simtab}).}

\item{As highlighted by \pcite{warren06}, the mass of halos determined using the FoF algorithm suffer 
from a systematic over-estimation due the statistical noise associated 
with sampling the halo density field with a finite number of particles.
The precise form of this correction has to be checked on a case by case basis.
We confirm the results of \cite{warren06}, particularly testing them for large halo masses, 
that the mass correction for the FoF(0.2) halos follows Eq.~(\ref{eq:warrencorrection}).
}

\item{For masses $M > 10^{14} \Msun$, we find 
an excess in the abundance of massive halos with respect to 
those from the Hubble volume simulation \cite{HVS,HVS2}, 
once corrected for the different cosmology used, 
that is a few times the Poisson error (see Fig.\ref{fig:massres}).
As argued in Sec.\ref{Sec:transients},
we conclude that this difference can be largely explained by systematics due to transients 
affecting the Hubble volume simulation.}

\item{From an extensive study of error estimates we 
conclude that: 
(i) sample variance is significant only for halos of $M < 10^{14} \Msun$, 
and dominates over shot-noise for box-sizes $L_{box}< 1 \Gpc$. Consistently, 
Poisson shot-noise errors under-estimate the total error budget by factors $2-4$ at $10^{12-13} \Msun$ in these volumes.
(ii) Jack-knife re-sampling
is in general consistent with external estimators such as ensemble averages, and theoretical errors.
However it underestimates the total error budget for small box-sizes 
($< 500 \Mpc$) where sampling variance is more important due to the lack of long-wavelength 
modes and to the fact that jack-knife regions are not independent.
(iii) For all our runs the theoretical error estimate in Eq.~(\ref{eq:sigmaMF}) is in remarkably good agreement with
our external {\it sub-volumes} estimator that incorporates fluctuations due to long-wavelength modes in addition to sampling and shot-noise variance at the scale of the given box-size.}

\item{Existing analytic fits \cite{ST99,warren06},
accurately reproduce our N-body measurements at up to $10^{14} \Msun$, but
fail to reproduce the abundance of the most massive halos,
underestimating the mass function by $10\%$ ($30\%$) at $M =  3.16 \times 10^{14}\Msun$ ($10^{15}\Msun$).}

\item{The FoF halo mass function deviates from universality (or self-similarity).
In particular, the Sheth \& Tormen (1999) fit, if extrapolated to $z\ge 0$ assuming universality 
leads to an underestimation of the abundance of $30\%$, $20\%$ and $15\%$ at $z=0$, $0.5$, $1$ 
for fixed $\nu = \delta_c/\sigma \approx 3$ 
(corresponding to $M\sim 7\times 10^{14}\Msun$, $2.5\times 10^{14}\Msun$ and $8\times 10^{13}\Msun$ respectively,
see Fig.~\ref{fig:MFfit_highz}). This is due to some extent to the ST not being a very good fit at $z=0$. If we instead extrapolate our $z=0$ MICE best fit, we level of evolution in the amplitude of the halo mass function is $\sim 5\%$ ($10\%$) at $z=0.5$ ($z=1$).}

\item{We provide a new analytic fit, Eq.~(\ref{eq:micefit}), that reproduces N-body measurements over more than
5 orders of magnitude in mass, and follows its redshift evolution up to $z=1$, that is accurate to $2\%$. 
The new fit has the functional form of
\cite{warren06}, but with different parameters to account for the excess in the high mass tail
(see also Table~\ref{fittab}).}

\item{Systematic effects in the abundance of clusters can strongly bias 
dark-energy estimates. We estimate that medium depth surveys $z\le 1$, using SZ cluster detection,
could potentially bias 
the estimated value of the dark-energy equation of state, $w$, by as much as $50\%$.
This effect is however an upper limit to the amplitude of this systematic effect, and it drops quickly with depth.
For deep surveys $z \sim 1.5$, such as DES, the estimated bias is largely negligible.}

\end{itemize}

As discussed in Sec.~\ref{sec:introduction} there are various ways to define halos and their masses, but the 
most commonly employed algorithms are Friends-of-Friends (FoF) and Spherical-Overdensities (SO). From an observer 
view-point it is not clear that a single definition is optimal for all kinds of detection 
techniques (see \pcite{voit05} for a review). X-rays observations are easier and more robust in the inner center of 
clusters, that have higher density contrasts and are more relaxed.
Hence, it seems clear that spherical apertures are more appropriate in this case, although with threshold
density contrasts as high as $500$ or more \cite{vikhlinin09}.
Optical richness is measured within regions of fixed projected physical radius, resembling a modified FoF method.
Hence, they demand a proper calibration of the mass-richness relation
\cite{bode01} if one does not want to rely on the correlation with the
cluster X-ray properties. Cluster detection through the Sunyaev-Zeldovich (SZ) effect are relatively recent in statistical
terms and suffers from several other sources of error (e. g.\pcite{carlstrom02}). 
Yet, both halo definitions have been employed in the literature. For
instance, mock SZ maps have been built and exploited 
using FoF \cite{schulz03}, and correlations between X-ray properties and the SZ signal were studied starting from SO catalogues \cite{stanek09}.

Our particular work focused on halos defined through the FoF percolation method leaving the SO for a follow-up study. Nonetheless we have tested that the trend seen towards high masses when comparing different MICE runs persist
for decreasing linking length, that is, when probing the inner most regions of the halos. 

We also leave for future work a more thorough investigation of the cosmological implications of our results.
In particular, it remains to be seen what is the impact of the bias on cosmological parameters induced by mass function priors 
in the presence of other systematic effects such as the uncertainties in the cluster mass-observable relations and 
the cosmological parameter degeneracies present in cluster abundance measurements.

\section*{Acknowledgments}
We would like to thank Jorge Carretero, Gus Evrard,  Marc Manera, Sebasti\'an Pueblas, 
Rom\'an Scoccimarro, Volker Springel, and Masahiro Takada 
for very useful help and discussions at different parts of this project. 
We acknowledge support from the MareNostrum supercomputer (BSC-CNS, www.bsc.es), through grants 
AECT-2008-1-0009, AECT-2008-2-0011, AECT-2008-3-0010,
and Port d'Informaci\'o Cient\'ifica (PIC, www.pic.es) where the simulations were ran and stored, 
the use of the Friends-of-Friends code from the University of Washington N-body shop (www-hpcc.astro.washington.edu) 
and the 2LPT code for initial conditions (cosmo.nyu.edu/roman/2LPT). 
The MICE simulations were implemented using the Gadget-2 code (www.mpa-garching.mpg.de/gadget).
Funding for this project was partially provided by the Spanish Ministerio de Ciencia e Innovacion (MICINN),
projects 200850I176, AYA2006-06341, Consolider-Ingenio CSD2007-00060, research project 2005SGR00728
from Generalitat de Catalunya and the Juan de la Cierva MEC program.

\end{document}